\documentclass[a4paper,10pt]{article}
\pdfoutput=1 

\newcommand{\mra}{\mathrm{a}}
\newcommand{\mrh}{\mathrm{h}}
\newcommand{\mrH}{\mathrm{H}}
\newcommand{\mrO}{\mathrm{O}}
\newcommand{\mrv}{\mathrm{v}}
\newcommand{\mrM}{\mathrm{M}}
\newcommand{\mrN}{\mathrm{N}}
\newcommand{\mrD}{\mathrm{D}}
\newcommand{\mcO}{\mathcal{O}}
\newcommand{\mrV}{\mathrm{V}}
\newcommand{\mrZ}{\mathrm{Z}}
\newcommand{\mrW}{\mathrm{W}}
\newcommand{\mrs}{\mathrm{s}}
\newcommand{\mrt}{\mathrm{t}}
\newcommand{\mru}{\mathrm{u}}
\newcommand{\mrq}{\mathrm{q}}
\newcommand{\mrA}{\mathrm{A}}
\newcommand{\mrT}{\mathrm{T}}
\newcommand{\mrG}{\mathrm{G}}
\newcommand{\mrF}{\mathrm{F}}
\newcommand{\mrB}{\mathrm{B}}
\newcommand{\mrdim}{\mathrm{dim}}
\newtheorem{proposition}{Proposition}
\newcommand{\mrR}{\mathrm{R}}
\newcommand{\mrQ}{\mathrm{Q}}
\newcommand{\mrd}{\mathrm{d}}
\newcommand{\mrL}{\mathrm{L}}
\newcommand{\mrf}{\mathrm{f}}
\newcommand{\distB}{\mathrm{dist}_{\mrB}}

\newcommand{\mrC}{\mathrm{C}}
\newcommand{\mrS}{\mathrm{S}}
\newcommand{\eqn}[1]{Eq.(\ref{#1})}

\newcommand{\SMEFT}{\mathrm{\scriptscriptstyle{SMEFT}}}
\newcommand{\HEFT}{\mathrm{\scriptscriptstyle{HEFT}}}
\newcommand{\myLO}{\mathrm{\scriptscriptstyle{LO}}}
\newcommand{\myNLO}{\mathrm{\scriptscriptstyle{NLO}}}
\newcommand{\mySM}{\mathrm{\scriptscriptstyle{SM}}}
\newcommand{\myBSM}{\mathrm{\scriptscriptstyle{BSM}}}
\newcommand{\LEBSM}{\mathrm{\scriptscriptstyle{LEBSM}}}

\newcommand{\tbt}{\mathrm{\overline{t} t}}
\newcommand{\bbb}{\mathrm{\overline{b} b}}

\newcommand{\sPW}{{\scriptscriptstyle{\mrW}}}
\newcommand{\sPZ}{{\scriptscriptstyle{\mrZ}}}
\newcommand{\sPB}{{\scriptscriptstyle{\mrB}}}
\newcommand{\sPD}{{\scriptscriptstyle{\mrD}}}
\newcommand{\sPA}{{\scriptscriptstyle{\mrA}}}
\newcommand{\sPG}{{\scriptscriptstyle{\mrG}}}
\newcommand{\sPV}{{\scriptscriptstyle{\mrV}}}
\newcommand{\sPR}{{\scriptscriptstyle{\mrR}}}

\newcommand{\OSs}{\mathrm{\scriptscriptstyle{OS}}}
\newcommand{\bGam}{{\overline{\Gamma}}}
\newcommand{\bM}{{\overline{\mrM}}}

\usepackage{jheppub} 

\usepackage[T1]{fontenc} 
\usepackage[utf8]{inputenc}
\usepackage{soul}
\usepackage[dvipsnames]{xcolor}
\usepackage[left]{lineno}

\usepackage{amsmath}

\renewcommand*{\backrefalt}[4]{%
 \ifcase #1 %
   (not cited)
 \or
(p.~#2)%
 \else
(pp.~#2)%
 \fi%
}
\usepackage{hypernat} 

\title{Use and reuse of SMEFT}

\author[a]{André David}
\author[b,c]{Giampiero Passarino}

\affiliation[a]{EP Department, CERN, Switzerland}
\affiliation[b]{Dipartimento di Fisica Teorica, Universit\`a di Torino, Italy}
\affiliation[c]{INFN, Sezione di Torino, Italy}

\emailAdd{andre.david@cern.ch}
\emailAdd{giampiero@to.infn.it}

\abstract{
In this work we address three questions: can we successfully describe (observed) deviations from the standard model  
in the SMEFT language? Can we learn something about the underlying, beyond the standard model, physics using the SMEFT 
language? If no deviation is observed, how to proceed?
Given the myriad of viable BSM options with extended scalar sectors, we suggest a widespread use of SMEFT not just 
as a global fitting tool (that could miss out on deviations from extended scalar sectors) but also as a bookkeeping 
framework in which the results from SMEFT fits to individual observables are provided, reported, and archived in 
a consistent way. The compatibility of such individual results can then be assessed in the light of BSM models 
with extended scalar sectors.


}

\begin{document} 
\maketitle


\flushbottom


 \section{Introduction \label{Intro}}

The SMEFT~\cite{Passarino:2016pzb,Brivio:2017vri,Passarino:2019yjx} is a framework that consistently extends the 
standard model (SM) and allows to capture the effects of beyond-standard-model (BSM) physics in a reasonably general fashion.

In order to define the SM effective{-}field{-}theory (SMEFT) we start by considering a broader scenario: there 
is a ``standard''  theory, $X$, described by a Lagrangian based on a symmetry group G. The definition of the EFT 
extension of $X$ (say, XEFT) requires a circumstantial description for which we need to consider $X^{\prime}$, the 
ultraviolet (UV) completion of $X$ or the next theory in a tower of EFTs.

The parameters of the ``standard'' $X$ theory are always measured to within some error. Having uncertainties in the 
parameters leads to hypothesizing a higher structure where the SM Higgs boson mixes with additional scalars.
Given the most recent results~\cite{1798909,Sirunyan:2018hoz,Sirunyan:2018kst,Aaboud:2018urx,Aaboud:2018zhk} we have 
to admit that this amount of mixing is observed to be rather constrained, especially because data continue to push 
the Higgs couplings towards the SM-like limits. 

There are two main, non-exclusive, paths in going from $X$ to $X^{\prime}$:

\begin{enumerate}
\item $X^{\prime}$ is based on the group G and contains heavy degrees of freedom belonging to some representation of G.
\item $X^{\prime}$ is based on a larger group F, where G$\subset\,$F and $X^{\prime}$ must reduce to $X$ at low energies.
\end{enumerate}

An additional assumption is that there are no ``undiscovered'' degrees of freedom in $X^{\prime}$ that are both light 
and weakly-coupled.
We can say that there are four players in the game: the standard theory $X$, the corresponding EFT extension
$X$EFT, the beyond{-}standard theory $X^{\prime}$ and its low{-}energy limit LE$X^{\prime}$. 

Most of this work (moving from the results presented in 
ref.~\cite{Passarino:2019yjx},
in section~$18$ of ref.~\cite{Brivio:2019irc},
and in ref.~\cite{ADtalk})
will be devoted to discuss the connection between $XEFT$ and LE$X^{\prime}$.

When $X$ is the standard model, $G = SU(3) \otimes SU(2) \otimes U(1)$, the simplest examples of extensions are the 
SM singlet extension~\cite{Chalons:2016jeu} (SESM), the THDMs containing two scalar doublets~\cite{Yagyu:2012qp}, or a 
non-supersymmetric $SO(10)$~\cite{Altarelli:2013aqa} which breaks down to the SM through a chain of different 
intermediate groups. Another example is the so-called 331 model~\cite{Okada:2016whh}.
For the SM, the EFTs are further distinguished by the presence (or absence) of a Higgs doublet in the construction.
In the SMEFT, the EFT is constructed with an explicit Higgs doublet. This is in contrast with the 
HEFT~\cite{Brivio:2013pma,Buchalla:2015qju} (an electroweak chiral Lagrangian with a dominantly $0+$ scalar)
that does not include such a doublet, i.e.\ no special relationship is assumed between the Higgs scalar and the Goldstone 
fields (see section~2 of ref.~\cite{Cohen:2020xca}). 

Additional selection criteria can be introduced for the SMEFT~\cite{Einhorn:2013kja}, 
in particular that a basis should be chosen from among Potentially-Tree-Generated (PTG) operators (as compared to LG, 
Loop-Generated operators). 

After having discussed the definition of the SMEFT we will consider several aspects of its implementation, i.e.\
the observational and mathematical consistency of the SMEFT will be critically examined in the light of known 
(but often overlooked) theoretical results. 

An interesting question is: what is so special about the SM? There seems to be no good answer so far:
spontaneous breaking of the EW symmetry in the minimal way does not necessarily mean that $SU(2) \otimes U(1)$
describes the most fundamental theory. We could imagine a scenario where, for some reason, people would have chosen 
a different ``standard'' theory, say $SU(3) \otimes U(1)$ gauge theory of the electroweak 
interactions~\cite{Lee:1977qs,Okada:2016whh}; what would be the present situation? Something similar to what we had during
the LEP{-}LHC interregnum: one scalar boson discovered and few heavy states to be fitted, therefore an incomplete 
``standard'' theory and no EFT. 

\subsection{Implementation of the SMEFT \label{imple}}

An important aspect of the SMEFT is the so{-}called SMEFT representation (linear or quadratic). 

\subsubsection{SMEFT representations}

Given any amplitude $\mrA$ its EFT expansion can be written as

\begin{equation}
\label{LvsQ}
\mrA = \mrA^{(4)} + \frac{1}{\Lambda^2}\,\mrA^{(6)} + \dots
\end{equation}

\noindent
where $\Lambda$ is the heavy scale. Therefore, when squaring the amplitude, ``linear'' means including the interference between $\mrA^{(4)}$ and $\mrA^{(6)}$; ``quadratic'' currently means including the square of $\mrA^{(6)}$ instead of the inclusion of all terms of $\mcO(1/\Lambda^4)$, on top of the $\mrdim = 8$ terms in the expansion, i.e. $\mrA^{(8)}$.
The obvious criticism to this procedure is that one should not construct $\mrS\,$-matrix elements at $\mathcal{O}(1/\Lambda^4)$ using a
canonically{-}transformed Lagrangian truncated at 
$\mathcal{O}(1/\Lambda^2)$.
When including $\mathcal{O}(1/\Lambda^4)$ terms the canonical normalization procedure induces changes to the shape of differential distributions, not only to their integral, as is the case for the aforementioned truncated Lagrangian.

The new results of ref.~\cite{Hays:2020scx} put a new perspective on the ``linear vs. quadratic'' option, discussing
(for the first time) a comparison between partial $\mathcal{O}(1/\Lambda^4)$ and full $\mathcal{O}(1/\Lambda^4)$.  

\subsubsection{Canonical normalization}

By canonical normalization we mean the problem induced when, given an effective Lagrangian, we find that the kinetic 
terms have a non{-}canonical normalization. This fact does not represent a real problem, as long as we remember 
the correct treatment of sources in going from amputated Green’s functions to $\mrS\,$-matrix elements.
To give an example we consider the SMEFT Lagrangian (Warsaw basis~\cite{Grzadkowski:2010es}) written in the mass 
eigenbasis; there will be terms like

\begin{equation}
\label{canR}
\begin{split}
\mathcal{L} &= - \frac{1}{2}\,\bigl( 1 +
\frac{\mrv^2}{\Lambda^2}\,\delta\mrZ^{(6)}_{\mrh} +
\frac{\mrv^4}{\Lambda^4}\,\delta\mrZ^{(8)}_{\mrh}
\bigr)\,\partial_{\mu} \mrh\,\partial_{\mu} \mrh
\\
&+ \frac{1}{\Lambda^2}\,\bigr( \mra^{(6)}\,\mrM^3_{\sPW}\,
\mrh \mrZ_{\mu} \mrZ_{\mu} +\,\dots\,\bigr) +
\frac{1}{\Lambda^4}\,\sum_i\,\mra^8_i\,\mcO^{(8)}_i ,
\end{split}    
\end{equation}

\noindent
where $\mrv$ is the Higgs VEV and the $\mra^{6,8}$ are Wilson coefficients; furthermore the $\delta\mrZ$ factors depend on Wilson coefficients. Note that the $\mrdim = 8$ terms, including $\delta\mrZ^{(8)}_{\mrh}$, are not
yet available. Canonical normalization means redefining the $\mrh$ field

\begin{equation}
\label{hshift}
\mrh = \bigl[ 1 - \frac{1}{2}\,\frac{\mrv^2}{\Lambda^2}\,\delta\mrZ^{(6)} + \mcO(\mrv^4/\Lambda^4) \bigr]\,\hat{\mrh} .    
\end{equation}

\noindent
As a consequence the Lagrangian becomes

\begin{equation}
\label{Lshift}
{\hat{\mathcal{L}}} = - \frac{1}{2}\,
\partial_{\mu} {\hat{\mrh}}\,\partial_{\mu} {\hat{\mrh}} +
\mra^{(6)}\,\frac{\mrM^3_{\sPW}}{\Lambda^2}\,
\bigl[ 1 - \frac{1}{2}\,\frac{\mrv^2}{\Lambda^2}\,
\delta\mrZ^{(6)}_{\mrh} \bigr]\,
{\hat{\mrh}}\,\mrZ_{\mu} \mrZ_{\mu} + \,\dots
\end{equation}

\noindent
showing terms of $\mcO(1/\Lambda^4)$ which are neglected in the ``quadratic'' representation.

Our approach to canonical normalization is completed with a rescaling of the SM parameters so that the part of the Lagrangian quadratic in the fields is the SM Lagrangian; for instance we use

\begin{equation}
\label{CNpar}
\mrM_{\sPW} \to \mrM_{\sPW}\,\Bigl( 1 - \frac{g_6}{\sqrt{2}}\,\mra_{\phi\sPW} \Bigr).
\end{equation}

\subsubsection{Equivalent operators}

Actually, there could be more $\mcO(1/\Lambda^4)$
``missing'' terms. First of all we need to define equivalent operators~\cite{Arzt:1993gz}: from the point of view of the $\mrS\,$-matrix two operators are equivalent if (for simplicity we will consider the case of scalar fields)

\begin{equation}
\label{redop}
\mcO_i - \mcO_j =
\mrF(\phi)\,\frac{\delta{\mathcal{L}}}{\delta\phi} ,
\end{equation}

\noindent
and we have to decide which one is to be eliminated (the redundant one) in order to construct a basis.
Redundant operators are eliminated by a field redefinition; the corresponding shift in the Lagrangian will eliminate redundant operators leaving a, neglected, higher order compensation which becomes relevant when we want to compare the SMEFT in two different bases and in the ``quadratic'' representation.

Building any EFT means promoting a theory with a finite number of terms into an effective field theory with an infinite number of terms and in doing so it is important to establish its consistency order-by-order.

\subsubsection{A provisional summary of the SMEFT}

In this paper, SMEFT will be understood as the SM extension containing dimension $6$ terms in the so{-}called Warsaw 
basis~\cite{Grzadkowski:2010es}.
However, in our approach, we have rescaled the Wilson coefficients: in front of an operator $\mathcal{O}^{(k)}_i$ of dimension $k$ and containing $n$ fields we write

\begin{equation}
\label{Wredef}
g^{n-2}\,\frac{\mra^{(k)}_i}{\Lambda^{k-4}}
\end{equation}

\noindent
where $g$ is the $SU(2)$ coupling constant.
This rescaling is useful when discussing SMEFT at the one{-}loop level, as explained in ref.~\cite{Ghezzi:2015vva}.

In conclusion: the SMEFT framework is useful because one can set limits on the effective coefficients in a model-independent way. This is why the SMEFT in the bottom-up approach, going beyond a global fit, is so useful: we do not know what the tower of UV-complete theories is (or if it exists at all) but we can formulate the SMEFT and perform calculations with it without needing to know what happens at arbitrarily high scales.
On the other hand, interpreting such limits as bounds on UV models (BSM models) does require some assumptions on the UV dynamics.

Having defined SMEFT, we further note that most BSM scenarios have extended scalar sectors. 
The lack of direct discovery of BSM states suggests that the SM is ``isolated,'' including a small mixing between 
light and heavy scalars~\cite{Wells:2016luz}. 
In other words, no ``light'' BSM scalars have been found and the light Higgs couplings seem to be SM-like.
The small mixing scenario raises the following question: if there are more scalars then we have to conclude that there 
is a small mixing with any other scalar. 
This is no longer accidental but a systematic effect, a feature of nature as we presently understand it for which 
there is no clear theoretical motivation for.

Mixing is not a peculiarity of extended scalar sectors; for instance we could consider extensions of the SM with 
general new vector bosons. It is worth noting that there are classes of BSM models where mass mixing terms of SM 
and new vectors are explicitly forbidden; however, the general case includes interactions with the Higgs doublet that 
give rise to mass mixing of the $\mrZ$ and $\mrW$ bosons with the new vectors when the electroweak symmetry is broken.  

\subsubsection{The SMEFT beyond LO}

An additional comment is about LO SMEFT vs.\ the inclusion of SMEFT loops, sometimes called NLO SMEFT.
Here, by NLO SMEFT we mean the following:

\begin{itemize}
    \item SMEFT vertices inserted in tree-level SM diagrams,
    \item tree-level (SMEFT-induced) diagrams with a non-SM topology,
    \item SMEFT vertices inserted in one-loop SM diagrams, and
    \item one-loop (SMEFT-induced) non-SM diagrams.
\end{itemize}

\noindent
``NLO'' SMEFT provides the general framework for consistent calculations of higher orders and allows for global fits, 
superseding any ad-hoc variation of the SM parameters.
Ongoing and near-future experiments can achieve an estimated per mille accuracy on precision Higgs and EW observables, 
thus providing a window to indirectly explore the theory space of BSM physics. That is why ``NLO SMEFT'' is needed.
To summarize:
NLO results have already had an important impact on the SMEFT physics program. LEP constraints should not be interpreted 
to mean that effective SMEFT parameters should be set to zero in LHC analyses. It is important to preserve the original 
data, not just the interpretation results, as the estimate of the missing higher order terms can change over time, 
modifying the lessons drawn from the data and projected into the SMEFT. Considering projections for the precision 
to be reached, LO results for interpretations of the data in the SMEFT are challenged by consistency concerns and are 
not sufficient, if the cut off scale is in the few TeV range. The assignment of a theoretical error for SMEFT analyses 
(missing higher order uncertainty or MHOU) is always important.

There is a hierarchy in the MHOU, ranging from $1\%$ to $100\%$; for instance, LO PDFs and NLO PDFs in LO SMEFT give 
results differing by a large factor. 

We can say that there aspects of the problem which should be solved ``today'' but 
not at the price of forgetting aspects which will require our attention ``tomorrow''. For instance, while QCD corrections 
are dominant it would be inaccurate to say that EW corrections are negligible (i.e. well below $10\%$), see 
refs.~\cite{Cullen:2019nnr,Cullen:2020zof} for an example. Furthermore, there are QCD corrections in the SMEFT which 
are unrelated to the SM ones and can be sizeable~\cite{Gauld:2016kuu}.
\subsubsection{The SMEFT and renormalization}

Further comments are: ``renormalization'' of any EFT should make UV finite all off-shell Green functions, 
i.e.\ not only those relevant for a single process.
In order to make all, on-shell, $\mathrm{S}\,$-matrix elements finite, we have to introduce renormalization for 
fields($\Phi$) and parameters ($p$), i.e.

\begin{equation}
\label{SMEFTren}
\Phi = \mrZ_{\Phi}\,\Phi_{\sPR}, \qquad
p = \mrZ_{p}\,p_{\sPR},
\end{equation}

\noindent
where the $\mrZ$ factors must be expanded order-by-order in $1/\Lambda^2$.
The full renormalization program~\cite{Ghezzi:2015vva} includes a) construction of the self-energies and Dyson 
resummation of the propagators; b) construction of $3$ (and higher) point functions, check of their $\mrdim = 4$ 
finiteness and complete removal of the residual $\mrdim = 6$  UV divergences by mixing Wilson coefficients.

There is a deep connection between UV poles and symmetry of the Lagrangian~\cite{Buchalla:2019wsc}; when including 
$\mrdim = 8$ operators we should realize that the SMEFT is computationally more complex than Quantum Gravity.
To give an example we consider a Lagrangian containing scalar field and use the 
background{-}field{-}method~\cite{tHooft:1973bhk} where we split the fields into a classical and a quantum part. 
All one{-}loop diagrams are generated by the part of the Lagrangian which is quadratic in the quantum fluctuations,

\begin{equation}
\label{BFM}
\mathcal{L}_2 = - \frac{1}{2}\,\partial^{\mu} \phi\,
\partial_{\mu} \phi +
\phi\,\mrN^{\mu}\,\partial_{\mu} \phi +
\frac{1}{2}\,\phi\,\mrM\,\phi .
\end{equation}

\noindent
In principle the counter{-}Lagrangian contains $8$ terms but if we define $X = \mrM - \mrN^{\mu} \mrN_{\mu}$ we can see 
that $\mathcal{L}_2$ is invariant under the 't Hooft transformation (H)

\begin{equation}
\label{Htransf}
\phi^{\prime} = \phi + \lambda\,\phi , 
\quad
\mrN^{\prime}_{\mu} = \mrN_{\mu} - \partial_{\mu} \lambda +
\bigl[ \lambda\,,\,\mrN_{\nu} \bigr] ,
\quad
X^{\prime} = X + \bigl[ \lambda\,,\,X \bigr] ,
\end{equation}

\noindent
where $\lambda$ is an antisymmetric matrix. Therefore, $\Delta\mathcal{L}$ will also be invariant, reducing the number 
of independent counterterms to $2$. Any EFT containing $\mrdim = 6$ and $\mrdim = 8$ operators will have terms like

\begin{equation}
\label{matval}
\frac{1}{2}\,\partial_{\mu} \phi_i\,\mathrm{g}^{\mu\nu}_{i j}(\phi_c)\,
\partial_{\nu} \phi_j ,
\end{equation}

\noindent
with a matrix{-}valued metric tensor. To $\mathrm{g}$ there will correspond matrix{-}valued Riemann tensors, i.e. 
many more invariants for $\Delta\mathcal{L}$~\cite{tHooft:1974toh}. Note that if $\mathcal{L}$ is invariant under a 
group G then the relation between G{-}invariance and H{-}invariance is crucial in proving closure under renormalization 
(not the same as strict renormalizability).

Once again today's priority goes to QCD NLO SMEFT, including those corrections that are unrelated to the SM.
All relevant $\mrdim = 6$ operators must be included and not only a subset, 
because subsets are, in general, not closed under renormalization; Wilson coefficients mix, e.g.\
there is a mixing between $\mra_{q\sPB\sPW}$ and $\mra_{q\sPG}$ in $\mrZ \to {\overline{q}} q$.

\subsubsection{The SMEFT and scheme dependence}

Scheme dependence is present in the SM and in the SMEFT predictions; it comes from ``finite renormalization'', i.e.\ the 
choice of experimental input quantities. A given observable $\mcO$ will be written as

\begin{equation}
\mcO = \mcO_{\myLO} + \mcO_{\myNLO} , \quad
\mcO_i = \mcO^{(4)}_i + g_6\,\mcO^{(6)}_i +\,\dots
\end{equation}

\noindent
where

\begin{equation}
\mcO^{(4)}_i \equiv \mcO^{(4)}_i(g_{\sPR}\,,\,\{\mrM_{\sPR}\}) ,
\quad
\mcO^{(6)}_i \equiv \mcO^{(6}_i(g_{\sPR}\,,\,\{\mrM_{\sPR}\}\,,\,
\{\mra_{\sPR}\}) ,
\end{equation}

\noindent
where counterterms and mixing of Wilson coefficients have been introduced and UV poles removed.
By LO we mean the lowest order in perturbation theory where the SM(SMEFT) observable is computed, 
e.g. $\mcO(g)$ for $\mrh \to \bbb$ and $\mcO(g^3)$ for $\mrh \to \gamma \gamma$ in the SM.
The definition of NLO requires more attention; for instance, the QCD corrections to the gluon fusion process, 
$g g \to \mrh$, require more than the two-loop calculation and must include

\begin{equation}
g g \to \mrh g , \quad
g q \to \mrh q , \quad
{\overline{q}} q \to \mrh g .
\end{equation}

On{-}shell finite renormalization requires $\mrM^2_{i, OS}$ to be a zero
of the real part of the inverse propagator for particle $i$. Then

\begin{equation}
\mrM_{i, \mrR} = \mrM_{i, OS} + \frac{g^2_{\sPR}}{16\,\pi^2}\,\Bigl[
d\mrZ^{(4)}_{\mrM_i} + g_6\,d\mrZ^{(6)}_{\mrM_i} \Bigr] .
\end{equation}

\noindent
in the $\mrG_{\mrF}\,$-scheme we require

\begin{equation}
g_{\sPR} = g_{exp} + \frac{g^2_{exp}}{16\,\pi^2}\,\Bigl[
d\mrZ^{(4)}_g + g_6\,d\mrZ^{(6)}_g \Bigr] ,
\end{equation}

\noindent
where $g_{exp}$ will be expressed in terms of the Fermi coupling constant $\mrG_{\mrF}$. The expressions for the 
renormalized quantities are then replaced into $\mcO$, truncating in $g_{exp}$ and in $g_6$. A complete $\mcO(1/\Lambda^4)$ 
calculation requires to perform finite renormalization at $\mcO(g^2_6)$ in order to be consistent.
Well{-}known arguments on the running of $\alpha_{em}$ and of $\mrG_{\mrF}$ indicate that the preferred scheme is based 
on selecting $\{\mrG_{\mrF}\,,\,\mrM_{\sPW}\,,\,\mrM_{\sPZ}\}$. However, a consistency check is based on the choice
$\{\alpha_{em}\,,\,\mrG_{\mrF}\,,\,\mrM_{\sPZ}\}$ where $\mrM_{\sPW}$ can be predicted, giving

\begin{equation}
\mrM_{\sPW} = \mrM_{\sPW}\mid_{\mySM} + \frac{\alpha_{em}}{\pi}\,g_6\,\Delta_{\sPW} ,
\end{equation}

\noindent
where the SMEFT corrections contain $9$ PTG and $9$ LG Wilson coefficients.

Input parameter sets values are based on extraction of the the parameters from different experimental results using 
the SM and not the SMEFT. Therefore, there is a problem, not only for $\alpha_s$ (although dominant) but 
for $\alpha_s, \mrM_{\sPW}, \mrM_{\sPZ}$ etc. Finally, SMEFT is not (yet) included in the PDF parametrization and all 
EWPD (e.g. LEP) are SM{-}based (LEPEWWG fits) and QED/QCD deconvoluted.

\subsubsection{The SMEFT vs. BSM}

Once again, we are considering the following scenario~\cite{Passarino:2019yjx}:
\begin{itemize}
    \item the SM, valid for $E << \Lambda$,
    \item the corresponding EFT extension (say SMEFT), and
    \item the next SM (NSM), some UV completion of the SM (or the next theory in a tower of effective theories).
\end{itemize}
We are interested in the low-$E$ limit of the NSM beyond the tree-level approximation.
\subsubsection{The SMEFT, BSM models and heavy{-}light contributions}

At the one{-}loop level we obtain local and non{-}local terms~\cite{delAguila:2016zcb,Jiang:2018pbd,Passarino:2019yjx} 
corresponding to long distance propagation and hence to reliable, perturbative, predictions at low energy, as well as 
local effects which, by contrast, summarize the unknown effects from high energies.
Having both local and non-local terms corresponds to a full implementation of the (one-loop) EFT program,
including the logarithmic dependence upon the characteristic momentum transfer in the problem, see 
ref.~\cite{Donoghue:2017pgk}.
To summarize: loop diagrams with light external legs and heavy internal ones admit a local low{-}energy limit; diagrams 
with light external legs and mixed internal legs may show normal-threshold singularities in the low{-}energy region and 
yield inherently non-local parts.

Any EFT Lagrangian and the corresponding EFT amplitudes have a different interpretation: the Lagrangian is local 
(as it should), the amplitudes generate long-distance kinematic logarithms.

As an example we consider a scalar $3\,$-point function

\begin{equation}
\label{C0NLexa}
i\,\pi^2\,\mrC_0(m\,,\,\mrM\,,\,m) = \int d^{\mrd}q\,
\Bigl[ (q^2 + m^2)\,((q + p_1)^2 + \mrM^2)\,((q + p_1 + p_2)^2 + m^2) \Bigr]^{-1},
\end{equation}

\noindent
in the limit $\mrM \to \infty$. The result is

\begin{equation}
\label{NLexa}
\mrC_0(m\,,\,\mrM\,,\,m) \sim \frac{1}{\mrM^2}\,\Bigl[
1 + \ln\frac{\mrM^2}{m^2} - \beta\,\ln\frac{\beta + 1}{\beta - 1}\Bigr] + \mathcal{O}(1/\mrM^4), 
 \end{equation}
 
\noindent 
where, using the Feynman prescription, $\beta^2 = 1 + 4\,m^2/(P^2 - i\,0)$ and $P = p_1 + p_2$.
The result shows the normal threshold at $P^2 = - 4\,m^2$. This example should be compared with

\begin{equation}
\label{C0Lexa}
\begin{split}
i\,\pi^2\,\mrC_0(\mrM\,,\,\mrM\,,\,\mrM) &= \int d^{\mrd}q\,
\Bigl[ (q^2 + \mrM^2)\,((q + p_1)^2 + \mrM^2)\,((q + p_1 + p_2)^2 + \mrM^2) \Bigr]^{-1},
\\
\mrC_0 &= \frac{1}{\mrM^2} + \mathcal{O}(1/\mrM^4).
\end{split}
\end{equation}

To summarize: special attention is due for configurations where there is a hierarchy involving the heavy scale, 
the Mandelstam invariants describing the process, and the light masses, 

\begin{equation}
\label{NLh}
\Lambda ^2 >> \mrs_{ij\,\dots\,k} =
-\,(p_i + p_j +\,\dots\,+ p_k)^2 >
(m_1 + m_2 + \,\dots\, + m_n)^2.
\end{equation}

\noindent
with the presence of normal{-}threshold effects and, eventually of anomalous{-}thresholds effects~\cite{Passarino:2018wix}.

\section{Using SMEFT \label{use}}

There is a large variety of directions in the SMEFT that allows for it to be a proxy for BSM scenarios.
The question is: can any BSM model in nature be caught by using the SMEFT? Well, ``a large variety'' means that we 
expect to have enough directions in the SMEFT to fit nearly everything.
But the underlying assumptions such as one single scalar doublet and one single heavy scale could affect the interpretation.
Therefore, if the question is ``can any BSM be caught by using an EFT with many assumptions?'' the answer will be: 
observable by observable, yes.
But will that lead us to what nature has in store?
No, not necessarily. But it will help because of the sensitivity in individual observables.

It is possible that for the BSM model realized in nature the effects in the full set of observables used is such that 
the SMEFT result \textit{seems} to be null. This can come about via an averaging effect, with different observables 
pulling a Wilson coefficient in opposite directions.

\begin{proposition}
Experiments cannot generate processes and reconstruct simulated event samples in every single BSM framework. 
SMEFT d.o.f.\ (the Wilson coefficients $\mra_i$) are being tested by experiments, are being expanded and improved upon, 
and are rather comprehensive as to the types of BSM deformations they can encode. SMEFT d.o.f. can be used as a 
bookkeeping tool in exploring the likelihood function, $L$, for (sub)sets of observables~\cite{ADtalk}.
\end{proposition}

An example of the procedure is given in App.~A of ref.~\cite{pulls} where pulls are introduced which can be interpreted 
in terms of fit robustness, bias and coverage.
In our case we can make separate fits of data samples characterized by, precision EW data, Higgs boson production 
(LHC Run 1 and Run 2), $\mrV\mrV$ production at LHC etc. Each fit yields estimates for the Wilson coefficients; 
estimates on one fit can be used to constrain the remaining fits. Note however that some of the data samples come 
with caveats, e.g. the correct interpretation of the $\mrh \to \mrZ \gamma$ signal strength.

Furthermore, if present low{-}energy measurements are not sensitive to a subset of SMEFT operators, there would be a null result which could be interpreted as the impossibility of uncover the corresponding heavy sector while a new set of measurements could very well do it. It is important to be able to quantify the impact of a new measurement in the SMEFT parameter space without having to redo the full fit. Bayesian inference has been suggested in ref.~\cite{vanBeek:2019evb}.

An additional warning is that at high scales there are $\mrdim = 8$ parameters with a greater impact than $\mrdim = 6$ 
parameters.
Inference about $\mrdim = 6$ parameters will be different if $\mrdim = 8$ is neglected; at the very least one should 
treat them as nuisance parameters and profile or marginalise them so as to obtain a (truncation) 
uncertainty~\cite{Hays:2018zze}.

\subsection{SMEFT and kappa parameters \label{kappas}}

An alternative way of recording SMEFT fits has been introduced in ref.~\cite{Ghezzi:2015vva} where a connection between 
Wilson coefficients and
kappa{-}parameters~\cite{LHCHiggsCrossSectionWorkingGroup:2012nn} was suggested.

In the original kappa-framework we replace
$\mathcal{L}_{\mySM}(\{m\}\,,\,\{g\})$ with
$\mathcal{L}_{\mySM}(\{m\}\,,\,\{\kappa_g\,g\})$, where $\{m\}$ denotes the SM masses, $\{g\}$ the SM couplings 
and $\kappa_g$ are the scaling parameters. This is the framework used during Run 1 of LHC.

In the SMEFT approach we define amplitudes: at LO

\begin{equation}
\label{genkLO}
\mrA^{\myLO}_{\SMEFT} = \sum_{i=1,n}\,\mrA^{(i)}_{\mySM} + i\,g_6\,\mrA_c ,
\qquad
g_6 = 1/(\sqrt{2}\,\mrG_{\mrF}\,\Lambda^2) ,
\end{equation}

\noindent
where the $\mrA^{(i)}_{\mySM}$ are the SM (gauge{-}parameter independent) sub{-}amplitudes and $\mrA_c$ is the 
SMEFT ``contact'' amplitude. For instance, in $\mrh \to \gamma \gamma$, the SM sub{-}amplitudes are the ones due 
to top, bottom and bosonic loops (the latter including $\mrW, \phi$ and FP ghosts). When $\mrdim = 6$ operators 
are inserted (once) in loops we obtain

\begin{equation}
\label{genkNLO}
\mrA^{\myNLO}_{\SMEFT} = \sum_{i=1,n}\,\kappa_i\,\mrA^{(i)}_{\mySM} +
i\,g_6\,\mrA_c +
g_6\,\sum_{i=1,N}\,\mra_i\,\mrA^{(i)}_{nf} ,
\end{equation}

\noindent
where the $\mra_i$ are Wilson coefficients and the $\kappa_i$ are linear combinations of the Wilson coefficients. 
Furthermore, the amplitudes $\mrA^{(i)}_{nf}$ collect all loop contributions which do not factorize into the 
SM sub{-}amplitudes.
The simplest example is $\mrh \to \gamma_{\mu}(p_1) + \gamma_{\nu}(p_2)$. The amplitude becomes

\begin{equation}
\label{hAA}
\mrA^{\mu\nu}_{\mrh \gamma \gamma} = i\,\mathcal{A}_{\mrh \gamma \gamma}\,
\bigl( p^{\mu}_2\,p^{\nu}_1 - p_1 \cdot p_2\,\delta^{\mu\nu} \bigr).
\end{equation}

We introduce $g^2_{\mrF} = 4\,\sqrt{2}\,\mrG_{\mrF}\,\mrM^2_{\sPW}$ and obtain

\begin{equation}
\label{Ac}
\mathcal{A}_c =
g_{\mrF}\,\frac{\mrM^2_{\mrh}}{\mrM_{\sPW}}\,\mra_{\sPA\sPA},
\end{equation}

\noindent
where $\mra_{\sPA\sPA} = c^2_{\sPW}\,\mra_{\phi\sPB} +
s^2_{\sPW}\,\mra_{\phi\sPW} + c_{\sPW}\,s_{\sPW}\,\mra_{\phi\sPW\sPB}$ and $s_{\sPW}$ is the sine of the weak{-}mixing angle.
The kappa coefficients in the factorizable part of the amplitude can be written as

\begin{equation}
\label{kandr}
\kappa_i = 1  + g_6\,\Delta\kappa_i = \frac{g^3_{\mrF}\,s^2_{\sPW}}{8\,\pi^2}\,\rho_i, \quad
\rho_i= 1 + g_6\,\Delta\rho_i,
\end{equation}

\noindent
where the index $i$ runs over $\mrW$ loops (i.e. the bosonic part), top quark loops, and b quark loops.
The non{-}factorizable part of the amplitude depends on the following Wilson coefficients,

\begin{equation}
\label{adep}
\mra_{q\sPW\sPB}\,,\,\mra_{\sPA\sPA}\,,\,\mra_{\sPA\sPZ}\,,\,\mra_{\sPZ\sPZ} ,
\end{equation}

\noindent
where we have defined

\begin{equation}
\begin{split}
\mra_{\sPZ\sPZ} &= s^2_{\sPW}\,\mra_{\phi\sPB} + c^2_{\sPW}\,\mra_{\phi\sPW} - s_{\sPW} c_{\sPW}\,\mra_{\phi\sPW\sPB} .
\\
\mra_{\sPA\sPZ} &=  \Bigl( 2\,c^2_{\sPW} - 1 \Bigr)\,\mra_{\phi\sPW\sPB} + 2\,s_{\sPW} c_{\sPW}\,\Bigl(
\mra_{\phi\sPW} - \mra_{\phi\sPB} \Bigr) .
\end{split}
\end{equation}

In the factorizable part of the amplitude, adopting the PTG scenario, we only keep $\mra_{q\phi}$ and 
$\mra_{\phi\sPD}\,,\,\mra_{\phi\Box}$.
These results tell us that the kappa{-}factors can be introduced also at the loop level; they are combinations of 
Wilson coefficients but we have to extend the scheme with the inclusion of process dependent kappa{-}factors and 
non{-}factorizable contributions.

The kappa{-}parameters form hyperplanes in the space of Wilson coefficients; each kappa{-}plane describes 
(tangent)flat-directions while normal directions are blind and there are correlations among different processes.

The generalized kappa parameters have two labels referring to the (gauge{-}parameter independent) SM sub{-}amplitude 
and to the process. The SMEFT requires relations among the $\Delta\kappa$, e.g.

\begin{equation}
\label{dkcorr}
\begin{split}
\Delta\kappa^{\mrh \gamma \sPZ}_b - \Delta\kappa^{\mrh \gamma \sPZ}_t &=
\Delta\kappa^{\mrh \gamma \gamma}_b - \Delta\kappa^{\mrh \gamma \gamma}_t
\\
\bigl( \frac{3}{2} + 2\,c^2_{\sPW} \bigr)\,
\bigl( \Delta\kappa^{\mrh \gamma \gamma}_t - \Delta\kappa^{\mrh \gamma
\mrZ}_t
\bigr) &=
c^2_{\sPW}\,\Delta\kappa^{\mrh \gamma \sPZ}_{\sPW} +
\bigl( \frac{1}{2} + 3\,c^2_{\sPW} \bigr)\,\Delta\kappa^{\mrh \gamma
\gamma}_{\sPW},
\end{split}
\end{equation}

\noindent
where the labels $t, b$, and $\mrW$ refer to the top quark loop, etc.
Another interesting relation concerns the $\mrh \tbt$ vertex where we have

\begin{equation}
\mrV^{\SMEFT}_{\mrh \tbt} =
\mrV^{\mySM}_{\mrh \tbt}\,\Bigl\{ 1 +
g_6\,\Bigl[\Delta\kappa^{\gamma\gamma}_t +
\frac{1}{2}\,c^2_{\sPW}\,
\Delta\kappa^{\gamma\sPZ}_{\sPW} -
\frac{1}{2}\,(2 - s^2_{\sPW})\,
\Delta\kappa^{\gamma\gamma}_{\sPW} \Bigr] \Bigr\} .
\end{equation}

In LO SMEFT the contact amplitude is non-zero while $\kappa_t$, etc. are set to one.
If a deviation is measured it will reflect into some value for the Wilson coefficient controlling the LO SMEFT.
However, at NLO SMEFT $\kappa_i \not= 1$ and we get a degeneracy; i.e.\ the interpretation in terms of LO SMEFT 
and NLO SMEFT could be rather different.

Another way of describing the generalized kappa framework and its connection with the SMEFT is as follows: we can write 
down all amplitudes respecting the required symmetries and these amplitudes are in one-to-one correspondence with 
the operators of SMEFT.

Of course, as soon as we start discussing the $\mrZ \gamma$ decay, a question will arise: we need a more general 
classification of the Higgs decays according to kinematics. Otherwise we will end up in a situation where it is not 
clear whether some event is actually $\mrh \to 4\,$fermions or rather 
$\mrh \to 2\,$fermions {+} radiation~\cite{Passarino:2013nka}.
For recent results on the Higgs decay into $\mrZ \gamma$ see ref.~\cite{collaboration2020search} where, however the most 
relevant theoretical results are (apparently) not used, see 
refs.~\cite{Kachanovich:2020xyg,Passarino:2013nka,Abbasabadi:1996ze,Abbasabadi:2006dd,Dicus:2013ycd}.

\section{How to reuse SMEFT \label{reuse}}

The increasing interest in the SMEFT has led to the development of a wide spectrum of public codes which implement 
automatically different aspects of the SMEFT for phenomenological applications~\cite{Brivio:2019irc}.

The question is: what happens if we only do global SMEFT fits to data from a nature with more scalars?
We will present few examples of BSM models and discuss their low{-}energy limits 
(for a similar discussion see ref.~\cite{Cohen:2020xca}) and then we will introduce the concept
that the SMEFT should not be understood (only) as a global fitting tool but (also) as a bookkeeping framework in which the 
results from SMEFT fits to individual observables are provided, reported, and archived. At the end of this section we
will discuss statistical aspexts of the procedure.

To summarize the New Physics (NP) scenario: given a BSM model, we

\begin{itemize}
\item[-] compute all relevant observables in terms of the Lagrangian parameters,

\item[-] take into account loop effects and the renormalization procedure.
We should keep in mind that there are subtle points in on-shell vs. MS renormalization vs. gauge invariance: 
tadpoles matter, i.e.\ they only cancel in on-shell renormalization. When masses of heavy states and mixings are 
MS-renormalized there could be problems: i.e.\ the MS-renormalization of the mixing angles combined with the popular 
on-shell renormalization schemes gives rise to gauge-dependent results already at the one-loop level~\cite{Denner:2018opp}.

\item[-] Compare to experimental results, i.e. Observables $\to$ Likelihood.
\end{itemize}

Admittedly this is time consuming to do for each BSM model. 

\begin{itemize}
\item The SMEFT is a powerful tool to connect model-building to phenomenology without needing to fit hundreds of 
observables to data in each model.
\end{itemize}

Before discussing specific examples of BSM models we look for theoretical guidance.
For instance, assuming an unbroken custodial invariance as suggested by precision electroweak 
measurements~\cite{Low:2012rj}, 
implies $\rho_{\myLO} = \mrM^2_{\sPW}/(c^2_{\sPW}\,\mrM^2_{\sPZ}) = 1$~\cite{Einhorn:1981cy,Passarino:1990nu,Lynn:1990zk}.
Note that Higgs doublets generally respect the custodial symmetry, except for certain combinations of the doublets 
associated with complex parameters.
Models with Higgs triplets can violate the symmetry.
Custodial symmetry can also be violated by terms of higher dimension arising from physics at some higher scale.

There is a comment to be made: the key point is not $\rho_{\myLO} = 1$ but the UV finiteness of $\rho$.
Therefore, renormalization is required in those BSM models where $\rho$ is UV-divergent; there are more Lagrangian 
parameters and, as a consequence, more counterterms which can be used to cancel the UV pole in $\rho$.
After removing the divergences we are left with ``finite'' renormalization, i.e.\ a scheme connecting the 
renormalized parameters to an experimental input containing $\rho_{exp}$.
We could claim that it is not a ``natural'' solution but it remains a solution.

\subsection{BSM models}

The archetype of BSM models is the so{-}called singlet extension of the SM (SESM). Here we summarize the approach 
followed in~\cite{Boggia:2016asg}.
The only modification w.r.t.\ the SM is contained in the scalar potential

\begin{equation}
\label{SESMpot}
- \mu^2_2\,\Phi^{\dagger} \Phi
- \mu^2_1\,\chi^2
- \frac{1}{2}\,\lambda_2\,\bigl( \Phi^{\dagger} \Phi \bigr)^2
- \frac{1}{2}\,\lambda_1\,\chi^4
- \lambda_{12}\,\chi^2 \Phi^{\dagger} \Phi .
\end{equation}

$\Phi$ is a doublet containing $\mrh_2$, the custodial singlet in $2_{\mrL}\,\otimes\,2_{\mrR}$, while

\begin{equation}
\label{chidef}
\chi = \frac{1}{\sqrt{2}}\,( \mrh_1 + \mrv_s ).
\end{equation}

The mixing angle is defined by

\begin{equation}
\label{amix}
\mrh = \cos\alpha\,\mrh_2 - \sin\alpha\,\mrh_1, \qquad
\mrH = \sin\alpha\,\mrh_2 + \cos\alpha\,\mrh_1 ,
\end{equation}

\noindent
where $\mrh$ and $\mrH$ are the mass eigenstates. We go to the mass eigenbasis, select

\begin{equation}
\label{defL}
\Lambda = \mrM_s = \frac{1}{2}\,g\,\mrv_s,
\end{equation}

\noindent
take the limit $\Lambda \to \infty$ and eliminate $\lambda_2$. At the same time, $\lambda_1$ and $\lambda_{12}$ 
(the ``extra SM'' parameters) remain free parameters, 

\begin{equation}
\label{tdef}
\lambda_{1} = \mrt_{1}\,g^2, \qquad
\lambda_{12} = \mrt_{3}\,g^2 .
\end{equation}

In this way $\lambda_2$ is modified (w.r.t.\ the SM) as follows,

\begin{equation}
\label{lam2}
\lambda_2 =     \frac{1}{4}\,g^2\,\frac{\mrM^2_{\mrh}}{\mrM^2_{\sPW}} + g^2\,
\frac{\mrt^2_3}{\mrt_1} + \mathcal{O}(\mrM^{-2}_s),
\end{equation}

\noindent
where $\mrM_{\mrh}$ is the (bare) mass of the light Higgs boson and $\mrM_{\sPW}$ is the (bare) mass of the $\mrW$ boson. 
It is worth noting that we did not ``integrate out'' the heavy degree of freedom in the weak eigenbasis 
(unphysical fields), where we can construct a manifestly $SU(2)\,\otimes\, U(1)$ invariant low energy Lagrangian 
by integrating out the $\mrh_1$ field in the limit 
$\mu_1 \to \infty$ (see ref.~\cite{Buchalla:2016bse} for a discussion of this construction).
In our approach the integration is performed in the (physical) mass eigenbasis; this is analogous to what is done 
in~\cite{Jenkins:2017jig} when deriving the low{-}energy Effective Field Theory below the electroweak scale (LEFT).

Relevant in this context is the argument of ref.~\cite{Gorbahn:2015gxa} on extended scalar sectors and mixing 
(see also ref.~\cite{Brehmer:2015rna}): integration in the weak eigenbasis reproduces the effect of scalar mixing on 
interactions involving one Higgs scalar $\mrh$, but fails to do so for the case of two scalars $\mrh\mrh$. 
Indeed, when integrating out the field $\mrh_1$ we obtain an effective Lagrangian where only the wave function of 
the Higgs field is modified w.r.t.\ the SM.
After integrating out the heavy field $\mrH$ (in the mass eigenbasis) we obtain a Higgs{-}gauge interaction leading 
to a mismatch between the $\mrh \mrV \mrV$ and the $\mrh \mrh \mrV \mrV$ couplings (w.r.t.\ their SM values).
Furthermore, one of the operators in the Warsaw basis is $\mcO_{\phi} = (\Phi^{\dagger}\,\Phi)^3$, 
where $\Phi$ is the $SU(2)$ doublet. Since 

\begin{equation}
\label{PhiPhi}
\Phi^{\dagger}\,\Phi = \frac{1}{2}\,\Bigl[
(\mrh_2 + \sqrt{2}\,\mrv)^2 + \phi^0 \phi^0 + 2\,\phi^{+} \phi^{-} \Bigr],
\end{equation}

\noindent
there will be one Wilson coefficient, $\mra_{\phi}$ for the couplings $\mrh^6_2\,,\,\mrh_2\,(\phi^0)^4$, etc. 
When integrating out the $\mrh_1$ field in the weak eigenbasis (at LO) we obtain $\mra_{\phi} = 0$.
But when integrating out the $\mrH$ field in the mass eigenbasis we obtain different coefficients in front of 
polynomials of scalar fields, a fact which becomes relevant when comparing SMEFT and SESM at the NLO level.

The key difference is that the SMEFT utilizes the complete Higgs doublet as a building block. On the other hand, 
the low{-}energy limit of some BSM model can be such that the physical Higgs boson excitation and the Goldstone 
bosons are independent objects.  
Another way of describing this fact is to look at the tree{-}generated scalar potential in the SESM 
(in the unitary gauge), where we obtain

\begin{equation}
\label{SPSESM}
\begin{split}
U(\mrh) &= - \frac{1}{2}\,\mrM^2_{\mrh}\,
\Bigl( \mrh^2 + \frac{1}{2}\,\frac{g}{\mrM_{\sPW}}\,
\mrh^3 + \frac{1}{16}\,\frac{g^2}{\mrM^2_{\sPW}}\,
\mrh^4 \Bigr)
\\
{} &+ \frac{1}{8}\,g\,\frac{\mrt^2_3}{\mrt^2_1}\,
\frac{\mrM^2_{\mrh}}{\Lambda^2}\,\Bigl(
2\,\mrM_{\sPW}\,\mrh^3 + \frac{7}{4}\,g\,\mrh^4 +
\frac{1}{4}\,\frac{g^2}{\mrM_{\sPW}}\,\mrh^5
\Bigr) ,
\end{split}    
\end{equation}

\noindent
showing the effect of the portal interaction $\lambda_{12}$ ($\mrt_{1,3}$ are given in \eqn{tdef}).
The loop{-}generated (local) part of the potential will not be reproduced here and can be found in 
ref.~\cite{Boggia:2016asg}. The potential in \eqn{SPSESM} follows after canonical normalization. Any Lagrangian 
for low{-}energy theories requires canonical normalization; in the SESM only the $\mrh$ field requires a shift,

\begin{equation}
\label{SESMCN}
\mrh \to \mrZ_{\mrh}\,\mrh '
\qquad\mrZ_{\mrh} = 1 - \frac{g^2}{96\,\pi^2}\,
\frac{\mrM^2_{\sPW}}{\Lambda^2}\,\frac{\mrt^2_3}{\mrt^2_1}\,(\mrt_1 - \mrt_3)^2 .
\end{equation}

\noindent
Actually, we can introduce shifts also for $\mrM_{\sPW}$ and $\mrM_{\mrh}$ so that the bare mass terms for physical states 
are SM{-}like. The shift in $\mrM_{\mrh}$ gives the typical ``fine{-}tuning'' which is often present when we derive 
the small of a low state from some UV completion. For comparison, in the SMEFT the Higgs potential is

\begin{equation}
\label{SMEFTpot}
U_{\SMEFT} = U_{\mySM} + \frac{g_6}{\sqrt{2}}\,\Bigl[ 
\mra_{\phi}\,U^{(6)}_{1} +  
(4\,\mra_{\phi\sPW} - \mra_{\phi\sPD} + 4\,\mra_{\phi\Box} )\,U^{(6)}_{2} + 
(\mra_{\phi\sPD} - 4\,\mra_{\phi\Box} )\,U^{(6)}_{3} \Bigr] ,
\end{equation}

\noindent
where we have included the effect of canonical normalization.
The four components are given by 

\begin{equation}
\label{tcomp}
\begin{split}
U_{\mySM} &= - 2\,\frac{\mrM_{\sPW}}{g}\,\beta_{\mrh}\,\mrh
- \frac{1}{2}\,\beta_{\mrh}\,\mrh^2 - \frac{1}{2}\,\mrM^2_{\mrh}\,\mrh^2
\\
{} &- \frac{1}{4}\,g\,\frac{\mrM^2_{\mrh}}{\mrM_{\sPW}}\,\mrh^3
- \frac{1}{32}\,g^2\,\frac{\mrM^2_{\mrh}}{\mrM^2_{\sPW}}\,\mrh^4 ,
\\
U^{(6)}_{1} &=
2\,g\,\mrM_{\sPW}\,\mrh^3
+ \frac{3}{2}\,G^2\,\mrh^4
+ \frac{3}{8}\,\frac{g^3}{\mrM_{\sPW}}\,\mrh^5
+ \frac{1}{32}\,\frac{g^4}{\mrM^2_{\sPW}}\,\mrh^6 ,
\\
U^{(6)}_{2} &= 
- \frac{1}{8}\,\beta_{\mrh}\,\mrh^2
- \frac{1}{16}\,g\,\frac{\mrM^2_{\mrh}}{\mrM_{\sPW}}\,\mrh^3
- \frac{1}{64}\,g^2\,\frac{\mrM^2_{\mrh}}{\mrM62_{\sPW}}\,\mrh^4 ,
\\
U^{(6)}_{3} &=
- \frac{1}{4}\,\frac{g}{\mrM_{\sPW}}\,\mrh\,\partial_{\mu} \mrh\,\partial_{\mu} \mrh
- \frac{1}{16}\,\frac{g^2}{\mrM^2_{\sPW}}\,\mrh^2\,\partial_{\mu} \mrh\,\partial_{\mu} \mrh ,
\end{split}
\end{equation}

\noindent
where $\beta_{\mrh}$ is designed to cancel tadpoles, order{-}by{-}order in perturbation theory.

The behavior of the mixing angle $\alpha$ is not selected a priori but follows from the hierarchy of VEVs,

\begin{equation}
\label{sinabeh}
\sin\alpha = \frac{\mrt_3}{\mrt_1}\,\frac{\mrM_{\sPW}}{\mrM_s}\,\Biggl[
1 + \Biggl( \frac{\mrt_2}{\mrt_1} -
\frac{3}{2}\,\frac{\mrt^2_3}{\mrt^2_1}\,
\frac{\mrM^2_{\sPW}}{\mrM^2_s} \Biggr) \Biggr] + \mathcal{O}(\mrM^{-5}_s).
\end{equation}

\noindent
where we have introduced

\begin{equation}
\label{t2def}
\mrt_2 = \frac{1}{4}\,\frac{\mrM^2_{\mrh}}{\mrM^2_{\sPW}} + \frac{\mrt^2_3}{\mrt_1} ,
\end{equation}

\noindent 
and $\mrt_{1,3}$ are defined in \eqn{tdef}.
The SESM is the simplest BSM model were we can discuss the general strategy. Once again, the SMEFT limit of the 
SESM can be obtained by integrating out the heavy field $\chi$,  while retaining the doublet $\Phi$ even though 
$\chi$ and $\Phi$ are not the physical fields.
Is the mixing subleading and diagonalization not needed? To answer this question we underline that the key parameter 
is the portal interaction between the doublet and the singlet fields~\cite{Brehmer:2015rna}, i.e. $\lambda_{12}$.
Our result is $\sin\alpha \propto \mrM_{\sPW}/\mrM_s$, see also ref.~\cite{Gorbahn:2015gxa};
additional suppression of the heavy mode can be obtained by requiring 
that $\lambda_{12} \propto g^2\,\mrM_{\sPW}/\mrM_s$~\cite{Walker:2013hka} (the so{-}called decoupling limit).
Only in this case are mixing effects moved to higher-dimensional operators (higher than $\mrdim = 6$).
It is worth noting that the SM decoupling limit cannot be obtained by making assumptions about only one parameter, 
indeed the relevant expansion parameters are $\lambda_{12}/\lambda_1$ and $\mrM_{\sPW}/\mrM_s$. We have adopted the 
more conservative approach, considering the non-decoupling limit where we keep $\lambda_{12}$ and $\lambda_1$ as 
free parameters of the effective theory.
In other words, the only assumption that we make is that the ratio of couplings is of the order of a perturbative 
coupling, i.e.\ $\lambda_{12}/\lambda^2_1 < 1/2$.

\paragraph{Summary of the SESM} \hspace{0pt} \\
in order to achieve $\mrM_{\mrH} \to \infty$ and $\mrM_{\mrh}$ finite with a mixing angle $\alpha \to 0$, one can consider 
the limit $\mu_1 \to \infty$ ($\mu_1$ being the only new mass scale in the singlet sector). However, 
keeping $\mrM_{\sPW}$ finite requires yet another modification. An obvious way out is provided by forcing decoupling 
of the singlet sector by taking $\lambda_{12} \to 0$ as 
well~\footnote{We acknowledge an important discussion with S.~Dittmaier}.
It is important to realize that the decoupling theorem (the Appelquist Carrazone theorem~\cite{Appelquist:1974tg}) 
tell us that the effects of heavy particles go into local terms in a field theory, either renormalizable  
couplings or in non-renomalizable effective interactions suppressed by powers of the heavy mass. 
In the SESM (as in many other models) we should replace the last statement with ``suppressed by powers of the heavy 
mass and/or by powers of the portal interaction''. 

The non{-}decoupling assumption does not exclude low values of $\lambda_{12}$ for which the effect of mixing becomes 
less relevant. For unitarity constraints on the Higgs portal see 
refs.~\cite{Walker:2013hka,Robens:2015gla,PhysRevD.88.115012}. One of the first examples of adding a scalar singlet 
in a gauge invariant way to the Higgs system can be found in ref.~\cite{Veltman:1989vw}, where renormalization is 
also discussed.

In the singlet extension of the SM we will have

\begin{equation}
\label{defPhi}
\Phi_{SESM} = \bigl[ \Phi_{SM(\mrh_2 \to \mrh)} \bigr] + \frac{1}{\sqrt{2}}\,
\bigl[ (\cos\alpha - 1)\,\mrh + \sin\alpha\,\mrH \bigr]\,e ,
\end{equation}

\noindent
where $e^{\dagger} = (1\,,\,0)$. 
It is worth noting that $\Lambda \not=\mrM_{\mrH}$ and that the limit $\Lambda \to \infty$ should be computed in 
the mass eigenbasis, not in the weak eigenbasis.
As a result, $\alpha$ is a function of $\Lambda$ and must also be expanded in powers of $\Lambda$.
This means that parts of the linear multiplets are integrated out, while other states are retained.
A large number of $1/\Lambda^2$ terms come from the expansion of the mixing angle, not from the integration of the 
heavy fields, unless $\lambda_{12}$ is further suppressed by an additional ``external'' choice or assumption. 
Therefore, we assume that the ratio $\mrt_3/\mrt_1$ of quartic couplings is $\mcO(1)$, without excluding strongly 
coupled scenarios~\cite{Brehmer:2015rna}.

As an example of mixing, we consider the operator
$\mcO_{weak} = {\widehat{\mcO}}\,(\mrD_{\mu}\Phi )^{\dagger} \mrD^{\mu} \Phi$ where ${\widehat{\mcO}}$ does not contain 
Higgs fields. We derive

\begin{equation}
\label{exmix}
\begin{split}
\mcO^{(\mrd + 4)}_{weak} &=
{\widehat{\mcO}}^{(\mrd)}\,\Bigl[
(\mrD_{\mu}\Phi_{\mrh})^{\dagger} \mrD^{\mu} \Phi_{\mrh}
- \sqrt{2}\,\sin^2\frac{\alpha}{2}\,g_{\mrh_2 \mrV \mrV}\,\mrh\,\mrV\,\mrV
\\
{} &+ \frac{1}{\sqrt{2}}\,\sin\alpha\,g_{\mrh_2 \mrV \mrV}\,\mrH \mrV \mrV
\Bigr] ,
\end{split}
\end{equation}

\noindent
where $g_{\mrh_2 \mrV \mrV}$ is the SM Higgs-$\mrV \mrV$ coupling and $\mrV = \mrW, \mrZ$. Furthermore, $\Phi_{\mrh}$ 
is the SM scalar doublet where $\mrh_2$ has been replaced by the light Higgs field in SESM. We have also made explicit 
the dimension of the operators.

\paragraph{THDM models} \hspace{0pt} \\
Another class of BSM models include additional doublets: the so{-}called THDM models.
There are four THDM models that differ in the fermion sector: they are type I, II, X and Y, see ref.~\cite{Yagyu:2012qp}.
The THDM models contain five physical states, two of which are neutral and even under CP transformations, one is 
neutral and CP-odd, and the remaining two carry the electric charge $\pm 1$ and are degenerate in mass.
It is assumed that the resonance measured at the LHC is the lighter CP-even Higgs $\mrh$, while the other particles 
are labelled $\mrH$, $\mrA$, and $\mrH^{\pm}$, respectively.
The above-mentioned THDM types contain eight independent parameters in the Higgs potential. 

Once again, the first problem in deriving the low{-}energy behavior of any BSM model is represented by the individuation 
of the cutoff scale; the SMEFT requires a unique scale, which implies a degeneracy of masses in the BSM model. Of course, 
multiple scales are relevant only if their effect is of the same order of $\mrdim = 8$ operators.
Two options have been discussed in the literature both in the unbroken phase \cite{Brehmer:2015rna} and in the mass 
eigenstates~\cite{Boggia:2016asg,Biekotter:2016ecg}, the latter based on the fact that custodial symmetry requires 
almost degenerate heavy states.

A more general situation would be the following: light masses ($m_i$) and two heavy masses 
with $m_i \ll \mrM_j$ but $\mid \mrM^2_1 - \mrM^2_2 \mid \ll \mrM^2_1 + \mrM^2_2$.
Given $\mrM^2_{-} = (\mrM^2_1 - \mrM^2_2)/2$, the most general result is given by a triple expansion, 
in $\mrM^{-2}_{1,2}$ and $\mrM^{-2}_{-}$, see refs.~\cite{Osipov:2001th,Passarino:2019yjx} for details.

\paragraph{SMEFT-predicted observables} \hspace{0pt} \\
Having discussed few examples of BSM models and their low{-}energy limits we are ready to formulate

\begin{proposition}{}
Can we successfully describe an observed SM{-}deviation in the SMEFT language?
Can we learn something about the underlying BSM physics using the SMEFT framework?
We can perform a fit of the SMEFT coefficients, $\{\mra\}$, to a set of observables, $\{\mrO\}$.
Take the best-fit results from the SMEFT fit to the data, $\{\hat{\mra}\}$, and compute ,SMEFT-predicted observables
$\{\hat{\mrO}\}_{\SMEFT} = \{\mrO\}_{\SMEFT}(\{\hat{\mra}\})$.
\begin{itemize}
\item Within framework $X$, with parameters $\vec{p}$, we compute $\{\mrO\}_{X}(\vec{p})$ and compare with 
$\{\hat{\mrO}\}_{\SMEFT}$.
\end{itemize}
There are two possible scenarios: we can perform the calculation directly using the Lagrangian of framework $X$ 
(for SESM see ref.~\cite{Altenkamp:2018bcs,Kanemura:2016lkz}) or we can compute the low{-}energy limit of $X$ and use 
the corresponding effective Lagrangian (for SESM see ref.~\cite{Boggia:2016asg}).

\end{proposition}

As discussed in section~\ref{kappas} there are ``flat directions'' in the space of Wilson coefficients; by this we mean 
that observables depend (at $\mrdim = 6$) on linear combinations of Wilson coefficients. This problem will show up 
whenever a limited set of experimental data is considered, e.g. $\mrN_{\mathrm{data}} < \mrN_{\sPW}$; of course, it is not
only a question of how many points but also what sensitivity the observables have to which operator.

A problem will remain if some of the linear combinations of the $\mra^6$ is poorly constrained by itself, a different
aspect of ``flat directions'' or ``sloppiness'' of the model.
For instance, consider vector{-}like fermions with opposite hypercharge: the operator $\mcO^{(6)}_{\sPW\sPB}$ is not
generated while the operator $\mcO_{\sPW}$ is LG, i.e.\ of $\mcO(g^3/16\,\pi^2)$. A way out could be to measure processes 
with an even number of $\mrB\,$-legs. Another example~\cite{Boughezal:2020uwq}, it is well known that the Higgs 
cross{-}sections alone cannot distinguish between Higgs couplings to gluons and top quarks.
Furthermore, consider the process ${\overline{\mru}} \mru \to \mrZ \mrh$ it depends on the following combinations
of Wilson coefficients

\begin{equation}
\label{Linc}
\begin{split}
\mra_{\mru\sPV} &= \mra_{\phi\mru} + \mra^{(1)}_{\phi\mrq} + \mra^{(3)}_{\phi\mrq} , \quad
\mra_{\mru\sPA} = \mra_{\phi\mru} - \mra^{(1)}_{\phi\mrq} - \mra^{(3)}_{\phi\mrq} , 
\\
\mra^{(1)}_n &= \mra_{\phi\sPD} - 4\,\mra_{\phi\sPW} , \quad
\mra^{(2)}_n = \mra_{\phi\sPD} + 4\,\mra_{\phi\sPW} + 4\,\mra_{\phi\Box}
\\
\mra_{\sPZ\sPZ} &= s^2_{\sPW}\,\mra_{\phi\sPB} + c^2_{\sPW}\,\mra_{\phi\sPW} - s_{\sPW} c_{\sPW}\,\mra_{\phi\sPW\sPB} .
\end{split}
\end{equation}

\noindent
For instance the ${\overline{\mru}} \mru \mrZ \mrh$ contact vertex is

\begin{equation}
i\,\frac{g g_6}{2 \sqrt{2} \mrM_{\sPW} c_{\sPW}}\,\gamma^{\mu}\,\Bigl( \mra_{\mru\sPV} - \mra_{\mru\sPA}\,\gamma^5 \Bigr).
\end{equation}

Alternative approaches have been proposed for dealing with the problem: diagonalization of the Fisher information 
matrix~\cite{Brehmer:2016nyr,MIG}
(FIM) or singular{-}value decomposition~\cite{Bodwin:2019ivc}. Sloppiness of a model is observed in many branches 
of physics. In those cases the Fisher information is ill{-}conditioned: one possibility is that the eigenvalues of 
the FIM are dependent on how the
model has been parametrized, i.e.\ we should choose a more natural parametrization from a phenomenological point
of view. In any case it is convenient to associate models with geometrical manifolds~\cite{TT}, with Wilson coefficients (or 
generalized kappas) as coordinates. 

To summarize: there are combinations of Wilson coefficients to which measurements are not sensitive. Technically
speaking the likelihood is flat (no curvature) when moving away from maximal likelihood. The results should be obtained from
a fit after re{-}parametrizing observables into a ``measurement'' recombination of Wilson coefficients. Different
operators cannot be disentangled by the measurement; threfore, only combinations are constrainable and it makes sense to use 
them as the new ``coordinates''.

All in all we insist on the fact that the main emphasis should be given to SMEFT{-}reconstructed observables and not to 
a list of Wilson coefficients, e.g. SMEFT observables can be reconstructed by fitting linear combinations of Wilson
coefficients. The problem becomes more complex when we include $\mcO(1/\Lambda^4)$ terms. There are two reasons for that,
there are $993$ $\mrdim = 8$ operators for one generation ($44807$ for $3$ generations and each observable depends on
linear combinations of $\mrdim= 6,8$ coefficients and on quadratic combinations od $\mrdim= 6$ coefficients.    
The latter originate from the square of $\mrdim = 6$ (one insertion) and from the interference between two $\mrdim = 6$
insertions and $\mrdim = 4$.
In conclusion it would be interesting to present tables similar to the ones produced at LEP with observables, 
measurements, SMEFT fits and pulls~\footnote{see for instance http://lepewwg.web.cern.ch/LEPEWWG/plots/winter2012/}.

Starting from any BSM Lagrangian we can compute observables, including one{-}loop diagrams and renormalization.
From any BSM model we can compute the low{-}energy limit, obtaining the corresponding effective Lagrangian, 
the LEBSM Lagrangian, which should be used consistently.
No additional problem will arise if we restrict the LEBSM to tree{-}generated operators. Special care must be 
adopted when loop{-}generated operators are included, as discussed in ref.~\cite{Boggia:2016asg}.
Consider, for example, the $\mrh \mrV \mrV$ vertex:
the tree{-}level generated vertex can be used in any LO/NLO calculation, i.e.\ it can be consistently inserted in 
one-loop diagrams containing light particles.
On the other hand, the loop{-}generated vertex can only be used, at tree level, in one loop calculations. 
I.e.\ it should not be inserted into loops of light particles.

One of the chief LHC statistical challenges is to devise techniques to test efficiently whether the data support 
the solid observation of an unexpected physics phenomenon or not.
For practical reasons, we may need a choice of the observables to consider.
This effectively means testing only a limited class of BSM extensions for which such choice shows enhanced sensitivity 
to processes in the kinematic range of the LHC.

Furthermore, we will have to check the statistical consistency between the simulated distribution of the BSM signal 
and the SMEFT signal~\cite{Kalinowski:2018oxd}.

Significance tests tell us how (statistically) confident we can be that there is truly a difference between a 
BSM model and the SMEFT.
For example: the null hypothesis, there is no ``real'' difference and the alternative hypothesis, there is a difference.
A significance test should measure how much evidence there is in favor of rejecting the null hypothesis.

\subsection{LHC, SM, BSM, EFT, distances and information}

When we work with a family of differential distributions (SM, SMEFT or BSM), there would seem to be an obvious way 
to proceed: calculate the distance between distributions. We ask how close we can come to guessing a BSM model, 
based on an observation.

We assume that a global SMEFT fit has been performed, returning the best value of the Wilson coefficients, 
$\{\hat{\mra}\}$ and their covariance matrix.
Consider a differential kinematic distribution, $\mrD(x)$; it could be $x = p_{\perp}$ and $\mrD = d\sigma/d p_{\perp}$ 
for a specific process; furthermore, $x \in X$. Given two distributions, $f(x)$ and $g(x)$ where

\begin{equation}
\label{norm}
\int_X dx\,f(x) = \int_X dx\,g(x) = 1 ,
\end{equation}

\noindent
we need to define their ``distance'', $\distB(f,g)$; one choice, used in the literature~\cite{Kalinowski:2018oxd},
would be the $\mrL^2\,$-norm of $f - g$. Here we prefer to use the so{-}called Bhattacharyya 
distance~\cite{Bhatta}, based on the following definitions:

\begin{equation}
\label{Bhatta}
\rho(f\,,\,g) = \int_{X} dx\,\sqrt{f\,g} ,
\qquad
\mathrm{dist}_{\sPB} = - \ln \bigl[ \rho(f\,,\,g) \bigr].
\end{equation}

Note that $X$ can be the full phase{-}space;
but instead of integrating over the entire phase space, it could be relevant to study how the information 
is distributed in phase space.

This distance satisfies $0 \le \rho(f\,,\,g) \le 1$ and $\rho(f\,,\,g) =
1$ iff $f = g$ while $\rho(f\,,\,g) = 0$ iff $f$ and $g$ are orthogonal.
If $X$ is split into a chosen number of bins then

\begin{equation}
\label{bins}
\rho(f\,,\,g) = \sum_{i=1}^n\, \sqrt{f_i\,g_i}
\end{equation}

\noindent
where $n$ is the number of bins and $f_i\,,\,g_i$ are the numbers of members of samples $f$ and $g$ in the $i\,$-th bin. 
The Bhattacharyya coefficient $\rho$ will be zero if there is a multiplication by zero in every bin.

The Bhattacharyya distance is related to the Hellinger distance (which obeys the triangular inequality) 
by $H^2 = 1 - \rho$. $H$ is a probabilistic analog of the Euclidean distance and can also be used to quantify the 
distance between measures from the same distribution indexed by different parameters,
$f(x\,;\,\theta_1\,,\,\dots\,,\theta_k)$~\cite{shemyakin2014}. This is particularly relevant when we want to compare 
a given differential distribution $\mrD$ in some BSM model with the corresponding one in the SM. Here, for example, 
$x = p_{\perp}$ and the $\theta$ parameters are the BSM ones, i.e. $\theta_i = 0, \forall i$ is the SM.

The concept of Hellinger information~\cite{shemyakin2014} is related to the Hellinger distance. Under certain 
regularity conditions it is closely related to Fisher information which has been shown to encode the maximum 
sensitivity of observables to model parameters for a given experiment~\cite{Brehmer:2017lrt,Brehmer:2016nyr}. 
Hellinger information can be also used to describe information properties of the parametric set in situations where 
the Fisher information does not exist. By means of the Hellinger distance we can obtain the robust estimators of 
multivariate location and covariance, as proposed in ref.~\cite{TamBoo}.
For additional usage of the Hellinger distance see ref.~\cite{Alvarez:2019knh,LOURENZUTTI20144414}.

The Fisher information metric~\cite{Brehmer:2017lrt} measures the amount of information a random variable $X$ contains 
in reference to an unknown parameter $\theta$.
The Fisher information distance is a consistent metric, enabling the approximation of the information distance when 
the specific parameterization of the manifold is unknown, and there have been many metrics developed for this approximation. 
The Hellinger distance is closely related to the information distance~\cite{carter2008information}.
Furthermore, Hellinger distance analogs of likelihood ratio tests have been proposed for parametric inference 
in ref.~\cite{10.2307/2289852}.
Finally, quoting ref.~\cite{canonne2018structure}, we can say that the classical Neyman{-}Pearson Lemma says 
that the ``standard'' test for distinguishing two distributions is the log{-}likelihood test but another classical 
test says that the optimal sample complexity is characterized by the square of the Hellinger distance.

Using a geometric framework we will discuss the interplay between SM, SMEFT and BSM models.
First of all we analyze the impact of the BSM signal: for that we maximize the distance $\distB(\mrD_{\myBSM},\mrD_{\mySM})$, 
varying the BSM parameters; this should be done under the condition that the BSM model remains a weakly coupled theory, 
e.g.\ the running coupling constants do not exceed some critical value and the conditions of vacuum stability
are satisfied for each value of the high scale $\Lambda$.
Therefore, conditions are necessary to single out parameter regions in the BSM model which cannot be treated 
perturbatively~\cite{Krauss:2017xpj}.
If the maximal distance is less than some, preselected, value then $\mrD(x)$ or $X$ are not a good choice.

Next we want to discuss BSM{-}SMEFT compatibility: for that we minimize the distance $\distB(\mrD_{\myBSM},\mrD_{\SMEFT})$ 
by varying both the BSM parameters and the Wilson coefficients under the following condition: we define a radius 
in the space of Wilson coefficients, $r^2 = \sum_i\,a^2_i$ and  require that $r \ge \hat{r}$ where 
${\hat r}^2 = \sum_i {\hat a}^2_i$. If the distance is greater than some, preselected, value then there will be a 
tension between the BSM model and the SMEFT.

At this point we can compare $\mrD_{\myBSM}$ and $\mrD_{\SMEFT}$ at the minimum of their distance with the band 
corresponding to $\mrD_{\SMEFT}$ reconstructed and derive informations on the goodness of the BSM model.

Finally, varying one SMEFT operator at a time is unlikely to be a useful description of UV-complete BSM physics.
For instance, without flavour assumptions, one needs to deal with a large number of independent operators corresponding 
to three fermion generations.
Because of this challenge, the complexity of the SMEFT analyses has, thus far, been restricted to a subset of 
higher-dimensional operators. More recently, a novel approach has been developed~\cite{vanBeek:2019evb}.

Multi-operator analysis and the combination of different observables is needed. The existence of additional operators 
in the HEFT{-}limit of BSM models may help.

\subsection{BSM contiguity \label{iso}}

The concept of SM isolation has been defined and discussed in ref.~\cite{Wells:2017aoy}.
Here we propose an alternative approach.
Within the $\mrdim = 6$ SMEFT approach we can only fit Wilson coefficients/$\Lambda^2$. For any BSM model the heavy scale is one of the parameters.
Let $\Lambda_{max}$ be the highest scale which can be tested at LHC. For a given differential distribution $D(x)$ we define
${\hat{\mrD}}_{\SMEFT}$ as the $\mrD\,$-distribution 
as the $\mrD\,$-distribution obtained by fitting the SMEFT to data. Given

\begin{equation}
\label{contdef}
\mathrm{dist}_{\myBSM} = \distB(\mrD_{\myBSM}\,,\,{\hat{\mrD}}_{\SMEFT}) ,
\end{equation}

\noindent
we minimize w.r.t. the BSM parameters (including the
heavy scale $\Lambda_{\myBSM}$, e.g. $\mrM_s$ in the SESM) Given a reference value for the distance, $\mrd_{ex}$ we have the following situations:

\begin{enumerate}

\item $\min\,\mathrm{dist}_{\sPB} > \mrd_{ex}$, the BSM model is excluded,

\item $\min\,\mathrm{dist}_{\sPB} < \mrd_{ex}$ and $\Lambda_{\myBSM} > \Lambda_{max}$. The BSM model and the SM are not contiguous (i.e. isolation of the SM).

\item $\min\,\mathrm{dist}_{\sPB} < \mrd_{ex}$ and $\Lambda_{\myBSM} < \Lambda_{max}$: the BSM model and the SM are contiguous.

\end{enumerate}

For a MC approach to the same problem see ref.~\cite{Hartland:2019bjb}.

\subsection{How good is the truncation error? \label{TU}}

Ref.~\cite{Biekotter:2016ecg} presents a discussion on the possible failure of $\mrdim = 6$ operators, in the low{-}energy limit of a BSM model (LEBSM), in describing LHC kinematics. This problem is equivalent to discussing the ``truncation error'' introduced in expanding observables in powers of $1/\Lambda$.

Given two distributions $f(x)$ and $g(x)$, with $x \in X$, their Hellinger distance can be written as

\begin{equation}
\label{Hdist}
H^2(f\,,\,g)= \frac{1}{2}\,\int_X dx\, \bigl(
\sqrt{f} - \sqrt{g} \bigr)^2.
\end{equation}

For a given process and a given distribution we want to compute the ``distance'' between $f = \mrD_{\myBSM}$, i.e. distribution $\mrD$ computed in the full BSM model, and its truncated, low{-}energy, expansion, i.e.

\begin{equation}
\label{leexp}
g = \mrD_{\mySM}\,\Bigl(
1 + \Delta_v\,\frac{\mrv^2}{\Lambda^2} + \Delta_e\,\frac{E^2}{\Lambda^2} \Bigr) ,
\end{equation}

\noindent
where $\Delta_v$ and $\Delta_e$ are, process 
dependent, kinematic factors and we have separated the ``scale''{-}growing contribution, i.e. $E$ can be any  scale describing the process while $\mrv$ is the Higgs VEV. We obtain that

\begin{equation}
\begin{split}
H^2\bigl( \mrD_{\myBSM}\,,\,\mrD_{\LEBSM} \bigr) &=
H^2\bigl( \mrD_{\myBSM}\,,\,\mrD_{\mySM} \bigr) 
\\
&- 
\int_X dx\,\bigl(
\sqrt{\mrD_{\myBSM}} - \sqrt{\mrD_{\mySM}} \bigr)\,\sqrt{\mrD_{\mySM}}
\Bigl(\Delta_v\,\frac{\mrv^2}{\Lambda^2} +
\Delta_e\,\frac{E^2}{\Lambda^2} \Bigr),
\end{split}
\end{equation}

\noindent
is as a quantity which can ``measure'' uncertainties associated to the truncation at $\mcO(1/\Lambda^2)$.

\subsection{SMEFT validity from unitarity bounds \label {uni}}

We should keep in mind that unitarity constraints must always be understood as ``perturbative unitarity''
constraints. Given a strictly renormalizable model depending on a parameter $p$, the statement $p < p_{max}$ means that
for $p > p_{\max}$ the model becomes strongly interacting. For the SMEFT we should distinguish between one{-}at{-}a{-}time
bounds and couple{-}channel bounds. For the sake of simplicity we consider the case of a single Wilson coefficient for which
we have derived an upper bound 

\begin{equation}
\mid \mrQ^2\,\frac{\mra^{(6)}_i}{\Lambda^2} \mid \;\le\; \mrA_i ,
\end{equation}

\noindent
where $\mrQ$ is the scale where the SMEFT is being tested. Suppose that, as a result of a fit, we have derived  
$\mid \mra^{(6)}_i/\Lambda^2 < \mrB_i \mid$. When the bound is saturated we conclude that
\begin{equation}
\mid \mrQ^2 \mid < \frac{\mrA_i}{\mrB_i} ,
\end{equation}

\noindent
or the SMEFT stops to be valid as a perturbative expansion.
These are not statements that unitarity is violated in the SMEFT. Unitarity ``would be violated'', if we could trust 
the perturbative expansion, which we cannot; there are perturbative unitarity bounds, but the bounds also imply that 
loops and higher dimensional operators must be important.
It is interesting to observe that there is complementarity between ($\mrh$) ``pole'' vs. ``tail'' measurements:
derivative operators influence tail observables and pole observables in a different way.
Tails are interesting, the accessible $\Lambda$ can be higher but, unfortunately, predictions will break in ``tails'' 
(or new physics will be seen before the breaking); projecting data into the SMEFT will have a large intrinsic 
uncertainty, i.e.\ we do not know what exactly is going on because the SMEFT interpretation becomes a series where 
the expansion parameter is close to $1$ and/or the perturbative unitarity bound is saturated.

UV complete models show then phenomenon of delayed unitarity~\cite{Ahn:1988fx} which is best seen in 
$\mrV \mrV\,$-scattering. If we have a 
light Higgs boson ($\mrh$) and an heavy one ($\mrH$), then the scattering could get strong for a range of energies, until
the high{-}energy UV physics starts unitarizing. The energy growing behavior is tamed only above $\mrM_{\mrH}$ and it is
expected if there is space enough between $\mrM_{\mrh}$ and $\mrM_{\mrH}$. 


\section{SMEFT vs. BSM models: a critical summary \label{csumm}}

There are several steps to be considered when comparing the SMEFT with a BSM model. First of all we recall the classification of ref.~\cite{Einhorn:2013kja}
where $\mrdim = 6$ operators can be PTG or LG; in any given BSM model, some operators may arise from tree diagrams, while others may only arise from loop corrections. The equivalence theorem~\cite{Kallosh:1972ap} relates some operators arising from loops to operators arising from trees; for the apparent puzzle of equivalence of $\mcO_{PTG}$, $\mcO_{LG}$
see section $4$ of ref.~\cite{Einhorn:2013kja}.
Imagine a situation where the SMEFT is defined by choosing as basis vectors PTG operators while the BSM model generates LG operators; this is exactly the scenario under discussion: the equivalence of two operators is a property of SMEFT while it is possible for the BSM theory to generate one but not the other. From this point of view we must be careful not to omit any basis operators, since their contributions to Green’s functions can be very different.

For the sake of simplicity we consider the decay $\mrh \to 4\,$leptons. We have the following situations:

\begin{itemize}

\item[1)] in the SMEFT we take the best available prediction for the $\mrdim = 4$ part~\cite{Denner:2019fcr} and add tree diagrams containing one $\mrdim = 6$ operator, see 
ref.~\cite{Brivio:2019myy}

\item[2)] In the BSM model we include tree diagrams and take the large $\Lambda$ limit (once $\Lambda$ has been identified).

\end{itemize}

\noindent
It follows that a comparison between 1) and 2) is not adequate since LG (local) operators have been included in the SMEFT. Therefore,

\begin{itemize}

\item[3)] we consider the BSM model at one{-}loop and take the large $\Lambda$ limit. However, in most cases, the result includes mixed heavy{-}light contributions which are not present in 1). As a consequence.

\item[4)] we include loops with one $\mrdim = 6$ operator insertion in the SMEFT predictions. Always following the PTG/LG classification, LG insertions call for a two{-}loop calculation in the BSM model.

\end{itemize}

There is an interesting connection between ultraviolet and infrared~\cite{Adams:2006sv,Zhang:2018shp}: the Froissart 
unitarity bound and dispersion relations imply a connection between unitarity in the UV (BSM models) and positivity in the
IR (SMEFT). The implication is that part of the observable parameter (Wilson coefficients) space is inconsistent with
causality and analyticity. 

\section{SMEFT and  LEP pseudo-observables \label{oldies}}

The main question is ``how to use (LEP 1) EWPD (POs) in the SMEFT analysis?''. In order to understand LEP 1 Pseudo{-}Observables (hadronic peak cross-section etc.) we have to clarify the strategy which was used~\cite{Bardin:1999gt}.

What the experimenters did~\footnote{Partially based on an old discussions with Manel~Martinez.} was just collapsing (and/or transforming) some ``primordial quantities'' (say number of observed events in some predefined set-up) into some ``secondary quantities'' or realistic{-}observables (RO) which are closer to the theoretical description of the phenomena.
In this step, if the number of quantities is reduced, this implies that some assumptions have been made on the behaviour of the primordial quantities. The validity of these assumptions is judged on statistical grounds. Within these assumptions (QED deconvolution,
resonance approach, etc.) the secondary quantities are as``observable'' as the first ones.
At this point, let us clarify that even the ``primordial quantities'', are obtained through 
many assumptions (event classification, detector response, etc.) which, as in the
previous case, can be judged just on statistical grounds.

The practical attitude of the experiments was to stay with a Model{-}Independent fit, i.e.  
from ROs $\to$ POs (plus a SM remnant) for each experiment, and these sets of POs were averaged.  
The result of this procedure are best values for POs. The extraction of Lagrangian
parameters was based on the LEP-averaged POs.

At LEP all QED initial state corrections and QED{+}QCD final state corrections were 
de-convoluted. The rationale for the de-convolution was based on the fact that all experiments used different kinematic cuts and selection criteria, while an objective
requirement was put forward by the scientific community for having universal
results anchored to the $\mrZ\,$-peak. Assuming a structure function representation for the initial fermions, in turn, allowed us to de-convolute the measurements and to access the hard scattering at the nominal peak.
Therefore, the transition from ROs to POs involves certain assumptions that reflect our understanding of SM effects.
In particular it was used the fact that in the SM there are several effects, such as the imaginary parts or $\mrZ{-}\gamma$ interference or the pure QED background, having a (usually) negligible influence on the line shape.
Therefore, POs are determined by fitting ROs but we will have some ingredients which are still taken from the SM, making the model{-}independent results (slightly) dependent upon the SM.
In other words, the information needed should contain a complete definition of lineshape and asymmetry POs, together with the residual SM dependence in model{-}independent fits; this includes a description on what is actually taken from the SM.

Another approach was also used at LEP, i.e. extraction of Lagrangian parameters directly from the ROs,  which are not raw data but rather educated manipulations of raw data, e.g. distributions defined for some simplified setup.

The main question to be answered is: are POs valid/usable even in the case where the SM is replaced by the SMEFT?
It is hard to believe that we will repeat the extraction of the SMEFT Lagrangian parameters (this time including Wilson coefficients) directly from the ROs. Therefore, the adopted strategy is to perform a SMEFT fit to the LEP POs;
guidance for estimating the corresponding uncertainty can be based on the following fact:
it has been tested by each LEP experiment how the results on the SM parameters differ between a SM fit to its own measured ROs, and a SM fit to the POs which themselves are derived in a fit to the same ROs.
For each experiment, the largest difference in central values, relative to the fitted errors, was observed for $\mrM_{\sPZ}$, up to $30\%$ of the fit error.
For the other four SM parameters, the observed differences in fitted central values and errors were usually below $10\% {-} 15\%$ of the fit error on the parameter.

Another effect has to do with scheme dependence. We should keep in mind that one of the key ingredients in computing the LEP POs has been $\alpha_{em}$ at the mass of the $\mrZ$. Define the runnnig of $\alpha_{em}$ as

\begin{equation}
\label{alphaatMZ}
\alpha_{em}(\mrM_{\sPZ}) = \alpha_{em}(0)\,\Bigl[
1 - \Delta\alpha^{(5)}(\mrM_{\sPZ})
- \Delta\alpha_{t}(\mrM_{\sPZ})
- \Delta\alpha^{\alpha\alpha_s}_{t}(\mrM_{\sPZ})  \Bigr]^{-1} ,
\end{equation}

\noindent
showing the top contribution, the mixed weak{-}QCD effects,
with $\Delta\alpha^{(5)}(\mrM_{\sPZ}) = \Delta\alpha_l(\mrM_{\sPZ}) +
\Delta\alpha^{(5)}_{had}(\mrM_{\sPZ})$, showing the leptonic part and the hadronic one. The SMEFT effect (neglecting LG operators in loops) is equivalent to replace

\begin{equation}
\label{SMEFTonalpha}
\Delta\alpha^{(5)}(\mrM_{\sPZ}) \to ( 1 - \kappa_{\alpha} )\,
\Delta\alpha^{(5)}_{\sPZ} ,
\end{equation}

\noindent
with $\kappa_{\alpha} = 0.188\,\mra_{\phi\sPD}$ at $\Lambda = 3\,$TeV, to give an example. As a result we obtain

\begin{equation}
\label{sizes}
\mid \kappa_{\alpha}\,\Delta\alpha_t \mid > \;\mid\; \Delta\alpha_l \mid ,
\quad
\mid \kappa_{\alpha}\,\Delta\alpha_t \mid \;\approx\;
\mid \Delta\alpha^{\alpha\alpha_s}_t \mid ,
\end{equation}

\noindent
i.e. there are SMEFT effects of the same order of magnitude than the SM $\mcO(\alpha_{em}\,\alpha_s)$
ones.

\section{Moving towards dimension \texorpdfstring{$8$}{8}}

There has been recent progress in building a $\mrdim = 8$ basis~\cite{Helset:2020yio,Murphy:2020rsh,Li:2020gnx}; imagine 
we use such a basis, say a generalized Warsaw basis, and that global fits have been performed giving the best values 
for the Wilson coefficients, $\{{\hat{\mra}}^{6,8}\}$. We recall a well{-}known result: removing a redundant operator
$\mcO^{(6)}_{\sPR}$ with coefficient $\mra^6_{\sPR}$ will propagate $\mra^6_{\sPR}$ into the Wilson coefficients 
of $\mrdim = 8$ operators~\cite{Passarino:2019yjx,Criado:2018sdb}. 

In the bottom{-}up approach the shift due to the field redefinition which eliminates $\mcO^{(6)}_{\sPR}$ can be 
absorbed into the coefficients of operators which are already present in the theory. Therefore, in this approach, 
we only ``measure'' combinations of Wilson coefficients, linear in $\{\mra^8\}$ and quadratic in $\{\mra^6\}$. 
To give an example we consider the operator

\begin{equation}
\mcO^{(6)}_{\sPR} = \Phi^{\dagger} \Phi\,\bigl(
\mrD_{\mu} \Phi \bigr)^{\dagger}\,
\mrD_{\mu} \Phi .
\end{equation}

\noindent
The term containing

\begin{equation}
g^2\,\frac{\mra^6_{\sPR}}{\Lambda^2}\,\mcO^{(6)}_{\sPR}
\end{equation}

\noindent
can be eliminated by the transformation~\cite{Passarino:2016saj}

\begin{equation}
\Phi \to \Phi - g^2\,\frac{\mra^6_{\sPR}}{\Lambda^2}\,
\bigl( \Phi^{\dagger} \Phi \bigr)\,\Phi ,
\end{equation}

\noindent
which induces higher order compensations, e.g.

\begin{equation}
\mcO^{(6)}_{\phi\sPB} \to
\mcO^{(6)}_{\phi\sPB} -
g^2\,\frac{\mra^6_{\sPR}}{\Lambda^2}\,
\mcO^{(8)}_{\mrh\sPB} ,
\end{equation}

\noindent
where the $\mrdim = 8$ operator is one of the operators of Tab.~$4$ in ref.~\cite{Hays:2018zze}.
However, consider some BSM model and compute its low{-}energy limit, obtaining some operator $\mcO^{(8)}_a$ with a 
coefficient $d_a$ which depends on the BSM parameters; in this limit we may also obtain some operator $\mcO^{(6)}_{\sPR}$
which is redundant in the basis where the fits have been performed and whose $\mrdim = 8$ compensation 
contains $\mcO^{(8)}_a$. As a consequence, when deriving relations between the set of Wilson coefficients $\{\mra^8\}$ 
and the BSM parameters, we should keep in mind that $d_a$ should not be compared with ${\hat{\mra}}^8_a$. Wrong 
relations should not be confused with ``systematic''. 

\section{Precision calls \label{Pcall}}

Precision calls~\footnote{Inspired by \url{https://en.wikipedia.org/wiki/Night\_Calls\_(album)}.}: 
the heavy particles in any BSM model are unstable and their description requires the introduction of the corresponding 
complex poles~\cite{Goria:2011wa}. The Dyson{-}resummed propagator for particle $\phi$ is
\begin{equation}
\label{Drp}
\Delta_{\phi}(s) = \Bigl[ s - \mrM^2_{\phi} + \Sigma_{\phi}(s) \Bigr]^{-1} ,
\end{equation}

\noindent
where $\mrM_{\phi}$ is the renormalized mass and $\Sigma_{\phi}$ is the renormalized $\phi$ self{-}energy (to all
orders but with one{-}particle{-}irreducible diagrams). The complex pole is defined as the complex solution of
\begin{equation}
\label{cpole}
s_{\phi} - \mrM^2_{\phi} + \Sigma_{\phi}(s_{\phi}) = 0 .
\end{equation}
To lowest order accuracy we can use
\begin{equation}
\Delta^{-1}_{\phi} = s - s_{\phi} ,
\end{equation}

\noindent
where the complex pole is conventionally parametrized as
\begin{equation}
\label{cppar}
s_{\phi} = \mu^2_{\phi} - i\,\gamma_{\phi}\,\mu_{\phi}.
\end{equation}
Let us define the following quantities:
\begin{equation}
\label{barred}
\bM^2_{\phi} = \mu^2_{\phi} + \gamma^2_{\phi} , \qquad
\mu_{\phi}\,\bGam_{\phi} = \bM_{\phi}\,\gamma_{\phi} .
\end{equation}

\noindent
It follows that
\begin{equation}
\label{barrap}
\frac{1}{s - s_{\phi}} = \Bigl(1 + i\,\frac{\bGam_{\phi}}{\bM_{\phi}} \Bigr)\,
\Bigr( s - \bM^2_{\phi}  + i\,\frac{\bGam_{\phi}}{\bM_{\phi}}\,s \Bigr)^{-1} ,
\end{equation}

\noindent
which is equivalent to say that we have introduced a running width with parameters which are not the on{-}shell ones.
Therefore, the low{-}energy limit of the propagator is controlled by barred parameters and not by the on{-}shell mass.
Let $\mrM_{\phi\OSs}$ and $\Gamma_{\phi\OSs}$ be the on{-}shell mass and width of the $\phi$, if
$\Gamma_{\phi\OSs}\,/\,\mrM_{\phi\OSs} << 1$ we can write a perturbative solution of \eqn{cpole},
\begin{equation}
\label{psol}
\mu^2_{\phi} = \mrM^2_{\phi\OSs} - \Gamma^2_{\phi\OSs} + \;\mbox{h.o.} \; ,
\qquad
\gamma_{\phi} = \Gamma_{\phi\OSs}\,\Bigl[ 1 - \frac{1}{2}\,\Bigl(
\frac{\Gamma_{\phi\OSs}}{\mrM_{\phi\OSs}} \Bigr)^2 \Bigr] + \;\mbox{h.o.}\;,
\end{equation}
 
\noindent
and the difference between barred and on{-}shell quantities is $\mcO(\Gamma^4_{\phi\OSs}\,/\,\mrM^4_{\phi\OSs})$. Outside this
region we have to solve \eqn{cpole} numerically.

To give an example we consider the SESM and introduce ${\overline{\mrt}}^2 = \mrt^2_3/\mrt_1$, so that
\begin{equation}
\sin\alpha \sim 2\,{\overline{\mrt}}\,\frac{\mrM_{\mrW}}{\mrM_\mrH} .
\end{equation}
Let $\Gamma_{\phi}(s)$ be the total width of particle $\phi$ at virtuality $s$; we obtain
\begin{equation}
\label{Hwidth}
\Gamma_{\mrH}(\mrM^2_{\mrH}) = \sin^2\alpha\,\Gamma^{\mySM}_{\mrh}(\mrM^2_{\mrH}) + \Gamma(\mrH \to \mrh \mrh) .
\end{equation}
The first component behaves like
\begin{equation}
\label{Fcomp}
4\,\bigl( {\overline{\mrt}}\,\frac{\mrM_{\mrW}}{\mrM_{\mrH}} \big)^2\,\Gamma^{\mySM}_{\mrh}(\mrM^2_{\mrH}) ,
\end{equation}

\noindent
where, for $\mrM_{\mrH} = 1\,$TeV the best calculation gives $\Gamma^{\mySM}_{\mrh} = 647\,$GeV. The second 
component~\cite{Bojarski:2015kra}, computed at LO, gives
\begin{equation}
\label{scomp}
\begin{split}
\Gamma(\mrH \to \mrh \mrh) &= \frac{K^2}{32\,\pi\,\mrM_{\mrH}}\,\Bigl(1 - 4\,\frac{\mrM^2_{\mrh}}{\mrM^2_{\mrH}}\Bigr)^{1/2} ,
\\
K &= g\,\sin\alpha\,\cos\alpha\,\Bigl(\mrM^2_{\mrh} + \frac{1}{2}\,\mrM^2_{\mrH}\Bigr)\,
\Bigl( \frac{\cos\alpha}{\mrM_{\mrW}} + 2\,\sqrt{\mrt_1}\,\frac{\sin\alpha}{\mrM_{\mrH}} \Bigr) .
\end{split}
\end{equation}

\noindent
Therefore, in the large $\mrM_{\mrH}$ limit we obtain
\begin{equation}
K \sim g\,{\overline{\mrt}}\,\mrM_{\mrH} ,
\qquad
\Gamma(\mrH \to \mrh \mrh) \sim \frac{g^2 {\overline{\mrt}}^2_1}{32\,\pi}\,\mrM_{\mrH} .
\end{equation}
For ``perturbative'' values of ${\overline{\mrt}}$ the ratio $\Gamma_{\mrH}/\mrM_{\mrH}$ remains small, the
difference between barred and on{-}shell parameters is negligible, i.e. $s/\mrM^2_{\mrH}$ and
$\Gamma_{\mrH}/\mrM_{\mrH}$ can be taken as the correct expansion parameters.

The advantage of \eqn{Hwidth} is that we can use the best available SM calculation fro$\Gamma^{\mySM}_{\mrh}$. However,
\eqn{Hwidth} is valid only at LO and , when loops are included , should be modified into
\begin{equation}
\label{mHwidth}
\Gamma_{\mrH}(\mrM^2_{\mrH}) = \sin^2\alpha\,\Gamma^{\mySM}_{\mrh}(\mrM^2_{\mrH}) + \Gamma(\mrH \to \mrh \mrh) +
\Delta\Gamma_{\mrH} .
\end{equation}
\noindent
The reason is that in $\Gamma^{\mySM}_{\mrh}$ we are using $\mrM_{\mrH}$ for Higgs couplings and Higgs propagators
(in loops). Let us define
\begin{itemize}

\item[-] $\Gamma^{s}_{\mrH}$ as the part of $\Gamma_{\mrH}$ containing loops with internal $\mrh$ and/or $\mrH$ lines,

\item[-] $\Gamma^{\mySM\,,\,s}_{\mrh}$ as the part of $\Gamma^{\mySM}_{\mrh}$ containing loops with internal $\mrh$ lines
of mass $\mrM_{\mrH}$.

\end{itemize}

\noindent
Then the additional contribution can be written as
\begin{equation}
\Delta\Gamma_{\mrH} = \Gamma^{s}_{\mrH} - \sin^2\alpha\,\Gamma^{\mySM\,,\,s}_{\mrh} .
\end{equation}

\noindent
It is worth noting that there are many different ingredients entering the calculation of $\Delta\Gamma_{\mrH}$, i.e.
loop diagrams, wave function factors, tadpoles and UV(finite) renormalization.

The presence of a resonance should be taken into account when considering the range of validity of the low{-}energy
expansion of any BSM model, i.e. we should not enter the region where the shape of the resonance is already visible.
Always using the SESM as an example we can ask for which values of $\mrM_{\mrH}$ the ratio $\Gamma_{\mrH}/\mrM_{\mrH}$ is
sizeable. For instance (at LO) we obtain
\begin{equation}
\Gamma(\mrH \to \mrZ \mrZ) \sim \sin^2\alpha\,\frac{\mrG_{\mrF}\,\mrM^3_{\mrH}}{16\,\sqrt{2}\,\pi} ,
\end{equation}
\noindent
and consider the ratio
\begin{equation}
x_{\sPZ} = \frac{\Gamma(\mrH \to \mrZ \mrZ)}{\mrM_{\mrH}} .
\end{equation}

\noindent 
To obtain $x_{\sPZ} = 1/2$ we need $\sin\alpha\,\mrM_{\mrH} = 1.74\,\mbox{TeV}$, which for $\sin\alpha = 0.2$ puts the
$\mrH$ resonance at $8.7\,$TeV. If we require $\sin\alpha = 0.2$ and $\mrM_{\mrH} = 3\,$TeV then
$\Gamma(\mrH \to \mrZ \mrZ) \approx 180\,$GeV and
$\Gamma(\mrH \to \mrW \mrW) \approx 360\,$GeV which gives a rough estimate of the scale where LESESM gives reliable
predictions.

\section{Examples \label{Exa}}

In this section we will discuss significant differences between SMEFT results and BSM models.
In order to discuss the low{-}energy limit of a BSM model, i.e. how the ``expansion'' is performed we
consider, once again, the following integral

\begin{equation}
\label{Idef}
\mathrm{I} = \mu^{\varepsilon}_{\sPR}\,\int d^{\mrd}q\,
\Bigl[(q^2 + \mrM^2_{\mrh})\,((q + p_1)^2 + \mrM^2_s)\,((q + p_1 + p_2)^2 + \mrM^2_{\mrh}) \Bigr]^{-1} ,
\end{equation}

\noindent
where $\mrM_s$ is a ``large'' scale and $\varepsilon = 4 - \mrd$; this integral will appear whenever there is a real, heavy, field $\mrS$ added to the SM Lagrangian, e.g. with an interaction $\Phi^{\dagger} \Phi \mrS$. To understand the details of the procedure we can say that there are $3$ ways of ``expanding'' the integral (see also ref.~\cite{Fuentes-Martin:2016uol}):

\begin{enumerate}
\item the heavy propagator is expanded as follows,

\begin{equation}
\label{expo}
\frac{1}{(q + p_1)^2 + \mrM^2_s} =
\frac{1}{\mrM^2_s}\,\Bigl( 1 - \frac{(q + p_1)^2}{\mrM^2_s} + \,\dots \Bigr)
\end{equation}    

\noindent
giving

\begin{equation}
\label{Iexpo}
\mathrm{I} \sim \frac{i \pi^2}{\mrM^2_s}\,\Bigl[
\frac{1}{\overline{\varepsilon}} - \ln\frac{\mrM^2_{\mrh}}{\mu^2_{\sPR}} + 2 -
\beta\,\ln\frac{\beta + 1}{\beta - 1} + \, \dots
\Bigr] ,
\end{equation}

\noindent
which corresponds to $\mrM^2_s >> \mid q^2 \mid \sim \mid p^2_i \mid$. This expression must be combined with EFT counterterms and it must be stressed that the soft part (the last two terms in the square bracket) cancels out in the matching procedure (although not a throwaway).

\item The heavy propagator is expanded while respecting the UV structure of the integral~\cite{vanderBij:1983bw} 
(at one loop)

\begin{equation}
\label{expt}
\frac{1}{(q + p_1)^2 + \mrM^2_s} =
\frac{1}{q^2 + \mrM^2_s}\,\Bigl( 1 -
\frac{p^2_1 + 2\,p_1 \cdot q}{q^2 + \mrM^2_s} + \,
\dots \Bigr)
\end{equation}

\noindent
giving the result of \eqn{NLexa}, 

\begin{equation}
\label{Iexpt}
\mathrm{I} \sim  \frac{i \pi^2}{\mrM^2_s}\,\Bigl[ 1 +
\ln\frac{\mrM^2_s}{\mrM^2_{\mrh}} -
\beta\,\ln\frac{\beta + 1}{\beta - 1} \Bigr] ,
\end{equation}

\noindent
corresponding to $\mrM^2_ s \sim \mid q^2 \mid >> \mid p^2_i$. This result includes both soft and hard terms.

\item The expansion can be performed \`a la Mellin{-}Barnes (instead of a Taylor expansion of the ``heavy'' propagators), see ref.\cite{Passarino:2019yjx} for details. Our findings are that the Mellin{-}Barnes expansion should always be the preferred method.

\end{enumerate}

It is worth noting that the expansion introduces kinetic logarithms. When we consider the addition of a complex, scalar, field to the SM and compute processes like $ g g \to \tbt$ the non{-}local terms will contain also di-logarithms. Obviously, heavy{-}light diagrams (above their normal threshold) develop an imaginary part leading to $\pi^2\,$-enhanced terms in the corresponding cross{-}section (another reason to include loops).

Most of our examples are dealing with the (real) singlet extension of the SM. The SESM Lagrangian contains several parameters. When taking the $\mrM_s = \Lambda \to \infty$ limit we have

\begin{equation}
\label{moredefS}
\sin\alpha \sim \mrf_2\,\frac{\mrM_{\sPW}}{\Lambda}
\quad\text{and}\quad
\mrM_{\mrH} \sim \mrf_1\,\Biggl( 1 + \mrf^2_2\,\frac{\mrM^2_{\sPW}}{\Lambda^2}\Biggr)\,
\Lambda,
\end{equation}

\noindent
where $\mrM_{\sPW}$ is the bare $\mrW$ mass and $\mrf_{1,2}$ are functions of the SESM parameters.

\subsection{\texorpdfstring{$\mrh \mrh \mrZ \mrZ$}{hhZZ}}

Let us consider the process
$\mrZ_{\mu}(p_3) + \mrZ_{\nu}(p_4) \to \mrh(p_1) + \mrh(p_2)$.
The LO SESM amplitude is the sum of $7$ diagrams and in the limit $\Lambda \to \infty$ we obtain

\begin{description}
\item [SMEFT]
\begin{equation}
\label{hhZZSMEFT}
\begin{split}
\mrA_{\mu\nu} &= 
\Biggl[ 1 + \frac{1}{3\,\sqrt{2}\,\mrG_{\mrF}\,\Lambda^2}\,
\Delta \Biggr]\,\mrA^{\mySM}_{\mu\nu}
\\
{} &+ \frac{1}{\sqrt{2}\,\mrG_{\mrF}\,\Lambda^2}\,
\frac{g^2}{c^2_{\sPW}}\,\bigl[
\mrF_1\,\delta_{\mu\nu} +
\mrF_2\,\mrT^{s\mrh}_{\mu\nu} +
\mrF_3\,\mrT^{s\mrh}_{\nu\mu} +
\mrF_4\,\mrT^{s\sPZ}_{\mu\nu} +
\mrF_5\,\mrT^{t}_{\mu\nu} +
\mrF_6\,\mrT^{u}_{\mu\nu}
\bigr]
\end{split}
\end{equation}

\item [SESM]
\begin{equation}
\label{hhZZSESM}
\begin{split}
\mrA_{\mu\nu} &= \Biggl( 1 - 2\, \lambda^2_2\,\frac{\mrM^2_{\sPW}}{\Lambda^2} \Biggr)\,
\mrA^{\mySM}_{\mu\nu} + \frac{g^2}{c^2_{\sPW}}\,\lambda^2_2\,
\frac{\mrM^2_{\sPW}}{\Lambda^2}\,\Biggl(
\frac{\mrT^{s\mrh}_{\mu\nu}}{\mrt - \mrM^2_{\sPZ}} +
\frac{\mrT^{s\mrh}_{\nu\mu}}{\mru - \mrM^2_{\sPZ}} \Biggr)
\end{split}
\end{equation}

\end{description}

\noindent
where we have introduced the following quantities:

\begin{equation}
\label{Tdef}
\begin{split}
\mrT^{s\mrh}_{\mu\nu} &= \mrM^2_{\sPZ}\,\delta_{\mu\nu} + p_{1\mu}\,p_{2\nu},
\\
\mrT^{s\sPZ}_{\mu\nu} &= p_{3\mu}\,p_{4\nu} +
\biggl(\frac{1}{2}\,\mrs - \mrM^2_{\sPZ}\biggr)\,\delta_{\mu\nu},
\\
\mrT^{t}_{\mu\nu} &= \bigl(\mrM^2_{\mrh} - \mrM^2_{\sPZ} - \mrt\bigr)\,
\delta_{\mu\nu} - p_{1\mu}\,p_{4\nu} - 
p_{3\mu}\,p_{2\nu},
\\
\mrT^{u}_{\mu\nu} &= \bigl(\mrM^2_{\mrh} - \mrM^2_{\sPZ} - \mru\bigr)\,
\delta_{\mu\nu} - p_{3\mu}\,p_{1\nu} - 
p_{2\mu}\,p_{4\nu},
\end{split}
\end{equation}

\begin{equation}
\label{Fdef}
\begin{split}
\Delta &= 6\,\mra_{\phi\sPW} - \mra_{\phi\sPD} + 10\,\mra_{\phi\Box},
\\
\mrF_1 &= 12\,\frac{\mrM^2_{\sPW}}{\mrs - \mrM^2_{\mrh}}\,\mra_{\phi} +
\frac{1}{4}\,\frac{\mrs}{\mrs - \mrM^2_{\mrh}}\,(
\mra_{\phi\sPD} - 4\,\mra_{\phi\Box} ) - \frac{1}{6}\,(
7\,\mra_{\phi\sPD} - 4\,\mra_{\phi\Box} ),
\\
\mrF_2 &= \frac{1}{6}\,\frac{1}{\mrt - \mrM^2_{\sPZ}}\,
(5\,\mra_{\phi\sPD} - 8\,\mra_{\phi\Box} ),
\\
\mrF_3 &= \frac{1}{6}\,\frac{1}{\mru - \mrM^2_{\sPZ}}\,
(5\,\mra_{\phi\sPD} - 8\,\mra_{\phi\Box} ),
\\
\mrF_4 &= \frac{1}{\mrM^2_{\sPZ}}\,\biggl(
3\,\frac{\mrM^2_{\mrh}}{\mrs - \mrM^2_{\mrh}} + 1 \biggr)\,
\mra_{\sPZ\sPZ},
\\
\mrF_5 &= \frac{2}{\mrt - \mrM^2_{\sPZ}}\,\mra_{\sPZ\sPZ}, 
\\
\mrF_6 &= \frac{2}{\mru - \mrM^2_{\sPZ}}\,\mra_{\sPZ\sPZ}.
\end{split}    
\end{equation}

Furthermore, $\mrs, \mrt$, and $\mru$ are Mandelstam invariants, $c_{\sPW}$ is the cosine of the weak{-}mixing angle, and $\mrG_{\mrF}$ is the Fermi coupling constant.

It is worth noting that in the SMEFT result we have made no distinction between PTG and LG operators.
We observe that the largest number of $1/\Lambda^2$ terms in SESM are due to the expansion of the mixing angle.
Furthermore, there are terms in SMEFT which are not reproduced by the SESM expansion, not even at higher orders in $1/\Lambda$.
We can twist SMEFT for this process but this would require (among other things) to set $\mra_{\sPZ\sPZ}$ to zero and this Wilson coefficient also multiplies the transverse part of the $\mrh \mrZ \mrZ$ LO vertex, contributing to SM{-}deviations in the decay $\mrh \to 4\,$-fermions.

The comparison has been performed using the LO SESM prediction. There will be many more terms when
SESM prediction is computed at the NLO level. The full SESM Lagrangian in the low{-}energy limit (the
LESESM Lagrangian) has been presented in ref.~\cite{Boggia:2016asg}. It is convenient to write a generic term in the LESESM Lagrangian as follows:

\begin{equation}
\label{gent}
\mcO = \frac{\mrM^l_{\sPW}}{\Lambda^n}\,{\overline{\psi}}^a\,\psi^b\,\partial^c\,(\Phi^{\dagger})^d\,\Phi^e\,\mrA^f,
\end{equation}

\noindent
where Lorentz, flavor and group indices have been suppressed,
$\psi$ stands for a generic fermion fields,
$\Phi$ for a generic scalar and $\mrA$ for a generic gauge field. All 
light masses are scaled in units of the (bare)
$\mrW$ mass. We define dimensions according to

\begin{equation}
\label{dimdef}
\mathrm{codim}\,\mcO = \frac{3}{2}\,(a + b) + c + d + e + f ,
\quad
\mathrm{dim}\,\mcO = \mathrm{codim}\,\mcO + l.
\end{equation}

Terms in the Lagrangian can be classified according to their dimension and their codimension. For instance, we obtain

\begin{equation}
\label{Ldc}
\begin{split}
\mathcal{L}_{6,6} &=
- \frac{1}{8}\,g^2\,\frac{\mrt^2_3}{\mrt^2_1}\,\,\partial_{\mu} 
\Phi^2_{\mrh}\,\partial_{\mu} \Phi^2_{\mrh}
- 
\frac{1}{384}\,\frac{g^4}{\pi^2}\,\frac{\mrt^2_3}{\mrt_1}\,\partial_{\mu} 
\Phi^2_{\mrh}\,\partial_{\mu} \Phi^2_{\mrh}
\\
&- 
\frac{1}{4096}\,\frac{g^6}{\pi^2}\,\frac{\mrt^3_3}{\mrt^3_1}\,\beta_1\,\Phi^6_{\mrh}
- \frac{1}{3072}\,\frac{g^6}{\pi^2}\,\frac{\mrt^3_3}{\mrt_1}\,\bigl(5 + 
9\,\mrA_0\bigr)\,\Phi^6_{\mrh}
\\
&+ 
\frac{3}{1024}\,\frac{g^6}{\pi^2}\,\frac{\mrt^3_3}{\mrt_1}\,\mrB_{00}\,\Phi^2_{\mrh} 
,
\end{split}
\end{equation}

\noindent
where $\Phi^2_{\mrh} = \mrh^2 + \phi^0 \phi^0 + 2\,\phi^+ \phi^-$. Furthermore, we have introduced

\begin{equation}
\label{ABdef}
\begin{split}
\mrA_0 &= \frac{2}{\mrd - 4} + \gamma + \ln\pi - 1 + 
\ln\frac{\mrM^2_{\mrH}}{\mu^2_{\sPR}} ,
\\
\mrB_{00} &= - \mrA_0 - 1 ,
\end{split}
\end{equation}

\noindent
where $\mrd$ is the space{-}time dimension, $\gamma$ is the  Euler{-}Mascheroni constant and $\mu_{\sPR}$ is the 't Hooft scale. The coefficient $\beta_1$ is due to tadpoles and is given by

\begin{equation}
\label{b1def}
\beta_1 = - 6\,\mrt^2_1\,\mrA_0.
\end{equation}

It is worth noting that the presence of a $\pi^2$ signals a loop{-}generated term.
The remaining components of the LESESM Lagrangian contain many more terms, for instance 
$\mathcal{L}_{6,5}$ contains

\begin{equation}
\label{Lsf}
- \frac{1}{512}\,\frac{g^5}{\pi^2}\,\frac{\mrt_3}{\mrt_1}\,\Bigl[ 6 + 
\bigl(9 + \frac{\mrt^2_3}{\mrt^2_1} \bigr)\,
\mrA_0 \Bigr]\,\Phi^2_{\mrh}\,\mrh\,\Bigl( 
2\,\mrM^2_{\sPW}\,\mrW^+_{\mu} \mrW^-_{\mu} +
\mrM^2_{\sPZ}\,\mrZ_{\mu} \mrZ_{\mu} \Bigr) ,
\end{equation}

\noindent
which is a loop{-}generated term.
Beyond the tree{-}generated terms the SESM Lagrangian is not as trivial as it may seem (simple
$\sin(\cos)\alpha\,$-rescaling of the SM results); for instance, consider the $\mrh^2 \mrH$ vertex

\begin{equation}
\label{notss}
- \frac{g}{2}\,\bigl( 2\,\mrM^2_{\mrh} + \mrM^2_{\mrH}
\bigr)\,\sin\alpha\,\cos\alpha\,
\Bigl( \frac{\cos\alpha}{\mrM_{\sPW}} + \frac{\sin\alpha}{\Lambda} \Bigr).
\end{equation}

In the limit $\Lambda \to \infty$ this vertex is not suppressed and that is why many loop diagrams containing heavy Higgs (internal) lines appear in the LESESM Lagrangian at $\mcO(1/\Lambda^2)$.

\subsection{\texorpdfstring{$\mrh \mrV \mrV$}{hVV}}

For instance, we can write

\begin{equation}
\label{hvv}
\mrh \to \mrV^{\mu}(p_1) + \mrV^{\nu}(p_2) =
\mrF^{\mrV\mrV}_{\sPD}\,\delta^{\mu\nu} + 
\mrF^{\mrV\mrV}_{\mrT}\,\mrT^{\mu\nu},
\end{equation}

\noindent
with $\mrT^{\mu\nu} = p_1\cdot p_2\,\delta^{\mu\nu} -
p^{\nu}_1\,p^{\mu}_2$.
Introduce $\mra_{\sPZ\sPZ} = s^2_{\sPW}\,\mra_{\phi\sPB} +
c^2_{\sPW}\,\mra_{\phi\sPW} - s_{\sPW}\,c_{\sPW}\,\mra_{\phi\sPW\sPB}$ and
$\rho = \mrM^2_{\sPW}/(c^2_{\sPW}\,\mrM^2_{\sPZ})$
and define

\begin{equation}
\label{Kdef}
\kappa^{\sPW\sPW} = - g\,\mrM_{\sPW}, \qquad
\kappa^{\sPZ\sPZ} = - g\,\frac{\mrM^2_{\sPZ}}{\mrM_{\sPW}}\,\rho
\end{equation}

\noindent
and also

\begin{equation}
\delta \kappa^{\sPW\sPW(\sPZ\sPZ)} =
\mra_{\phi\sPW} + \mra_{\phi\Box} \mp \frac{1}{4}\,\mra_{\phi\sPD}
\end{equation}

\noindent
to derive

\begin{equation}
\label{ehVV}
\begin{split}
\mrh \to \mrW^-_{\mu}(p_1) + \mrW^+_{\nu}(p_2) &=
\kappa^{\sPW\sPW}\,\Bigl(
1 + \frac{g_6}{\sqrt{2}}\,\delta\kappa^{\sPW\sPW} \Bigr)\,
\delta_{\mu\nu} - \sqrt{2}\,\frac{g\, g_6}{\mrM_{\sPW}}\,
\mra_{\phi\sPW}\,\mrT_{\mu\nu},
\\
\mrh \to \mrZ_{\mu}(p_1) + \mrZ_{\nu}(p_2) &=
\kappa^{\sPZ\sPZ}\,\Bigl(
1 + \frac{g_6}{\sqrt{2}}\,\delta\kappa^{\sPZ\sPZ} \Bigr)\,
\delta_{\mu\nu} - \sqrt{2}\,\frac{g\, g_6}{\mrM_{\sPW}}\,
\mra_{\sPZ\sPZ}\,\mrT_{\mu\nu}
\end{split}
\end{equation}

\noindent
where $\sqrt{2}\,g_6 = 1/(\mrG_{\mrF}\,\Lambda^2)$.
As a consequence, SMEFT predicts a change in the normalization of the SM-like term and the appearance of the transverse term. Note that $\mra_{\phi\sPD}$ induces a breaking of custodial symmetry.

This result should be compared with

\begin{equation}
\label{compare}
\Biggl( 1 + \frac{c^{\mrV\mrV}_1}{\Lambda^2}\,\mrh +
\frac{c^{\mrV\mrV}_2}{\Lambda^2}\,\mrh^2 + \dots\Biggr)\,
\mrV_{\mu}\,\mrV^{\mu} + \dots \in \mathcal{L}_{\HEFT},
\end{equation}

\noindent
or, more generally, with

\begin{equation}
\label{mgen}
\Biggl( 1 + \frac{d^{\mrV\mrV}_1}{\Lambda^2}\,\mrh +
\frac{d^{\mrV\mrV}_2}{\Lambda^2}\,\mrh^2 + \dots\Biggr)\,
\mrF^a_{\mu\nu}\,\mrF^{a\,\mu\nu} + \dots \in \mathcal{L}_{\HEFT},
\end{equation}

The $c^{\mrV\mrV}_{1}$ and $c^{\mrV\mrV}_{2}$ give information on the doublet structure of the scalar field and can be computed in any UV completion of the SM. For the $\mrh \mrV \mrV$ vertex in the SESM we will have SM{-}like terms of $\mcO(g)$, tree{-}generated terms of $\mcO(g/\Lambda^2)$ and loop{-}generated terms of $\mcO(g^3/\pi^2)$ containing both $\mcO(1)$ and $\mcO(1/\Lambda^2)$ components.

The SMEFT prediction is

\begin{equation}
\label{cratios}
\mrR_{\sPW} = \frac{c^{\sPW\sPW}_2}{c^{\sPW\sPW}_1} =
\frac{1}{2}\,\frac{g}{\mrM_{\sPW}}\,\Bigl(
 1 - \frac{g_6}{\sqrt{2}}\,\delta\kappa^{\sPW\sPW} \Bigr) ,
 \quad
 \mrR_{\sPZ} = \frac{c^{\sPZ\sPZ}_2}{c^{\sPZ\sPZ}_1} =
\frac{1}{2}\,\frac{g}{\mrM_{\sPW}}\,\Bigl[
 1 - \frac{g_6}{\sqrt{2}}\,(
 \delta\kappa^{\sPZ\sPZ} - \mra_{\phi\sPD} ) \Bigr] 
\end{equation}

\noindent
where $\delta\kappa^{\sPW\sPW} = \delta\kappa^{\sPZ\sPZ}$ if $\mra_{\phi\sPD} = 0$ (custodial symmetry).
If the fit to data yields $\mrR_{\sPW} \not= \mrR_{\sPZ}$, the SMEFT interpretation will be $\mra_{\phi\sPD} \not= 0$, though several new terms will appear if $\mrdim = 8$ operators are included.
It is worth noting that in the SESM expansion, $\mrF^{\mrV\mrV}_{\mrT}$ starts at $\mathcal{O}(1/\Lambda^4)$ and arises only from loops with both light and heavy internal propagators.
The general conclusion is that it is important to compare processes with one $\mrh$ leg and two $\mrh$ legs, e.g. the $\mrh$ decay into $4$ leptons vs.\ double Higgs production (in vector boson scattering).

There are subtle points in comparing the SMEFT
with BSM models; we can have SMEFT at LO, including both PTG and LG operators (vertices) and the SMEFT with loops where one $\mrdim = 6$ operator is inserted (eventually PTG only). At the same time we can have the full BSM model, its low{-}energy limit at the LO level, the same limit including loops. It is interesting to observe that, in the SESM, heavy{-}light contributions are $\mcO(1/\Lambda^4)$ for the $\mrh \mrV \mrV$ vertex but they are $\mcO(1/\Lambda^2)$ for the $\mrh \mrh \mrv \mrV$ vertex. 

\subsection{\texorpdfstring{$\mrh\mrZ$}{hZ} production}

As a final example we consider the $p_{\perp}$ distribution of the $\mrZ$ boson in the process ${\overline{q}} q \to \mrZ \mrh$.
In SESM, at LO there is a simple rescaling of the SM predictions.
In SMEFT (at LO) we have an expression containing $9$ Wilson coefficients, $3$ of them loop generated (the full SMEFT amplitude has been presented in section~6.3 of ref.~\cite{Passarino:2019yjx}).

There are two possible scenarios:
scenario a)
it is possible to obtain an almost constant SMEFT/SM ratio with special values of the Wilson coefficients, e.g.\ $\mra_{\sPA\sPZ} = - 1/(16\,\pi^2)$ and other coefficients set to zero.

Scenario b) shows a completely different behavior. For instance, with $\mra_{\phi\mrd\mrV} = 1$ and other coefficients set to $-1$ we obtain that the SMEFT(linear)/SM ratio becomes negative at around $350$ GeV, while the SMEFT(quadratic)/SM ratio remains positive, but exploding for increasing values of $p_{\perp}$.
For $p_{\perp} \approx 500\,$GeV the ratios are $-\,2$ for the linear representation and $+\,4$ for the quadratic one, an evident sign of the breakdown of the SMEFT.

The SESM option (and many others) is excluded if scenario b) is the result of the SMEFT fit. 

\subsection{\texorpdfstring{${\overline{t}}\,t\,\mrh$}{tth}}

Here we consider $g g \to \tbt \mrh$. In the SMEFT we will have to assemble several vertices; as a result we will have factorizable (fct) and non{-}factorizable (nfct) contributions, the latter changing the shape of distributions. For instance, the $g \tbt$ vertex will become

\begin{equation}
\label{fandnf}
\begin{split}
\mrV^{\mu\,,\,a}_{i j}\,\mid_{fct} &= \frac{i}{2}\,g_{\mrs}\,\gamma^{\mu}\,
\phi\lambda^{a}_{i j}\,
\Bigl(1 + \frac{g_6}{\sqrt{2}}\,\mra_{\phi \mrG}\Bigr) ,
\\
\mrV^{\mu\,,\,a}_{i j}\,\mid_{nfct} &= -
\frac{i}{4}\,g_{\mrs}\,g_6\,\mra_{\mrt\mrG}
\lambda^a_{i j}\,\sigma^{\mu\nu}\,p_{\nu} ,
\end{split}
\end{equation}

\noindent
where $p$ is the (ingoing) gluon momentum.
Next we will have the (factorizable) $\mrh \tbt\,$vertex,

\begin{equation}
\label{factV}
\mrV = - \frac{1}{2}\,\frac{m_t}{M_\sPW}\,\Bigl[
  1 + \frac{g_6}{\sqrt{2}}\,\bigl(
\mra_{\phi \sPW} + \mra_{t \phi} - \frac{1}{4}\,\mra_{\phi \sPD} +
\mra_{\phi \Box} \bigr) \Bigr],
\end{equation}

\noindent
the (factorizable) $3\,$-gluon vertex which receives a correction $g_6/\sqrt{2}\,\mra_{\phi \mrG}$.
Here $g_{\mrs}$ is the strong coupling constant. Furthermore,

\begin{equation}
\label{convent}
\mathrm{Tr}\,(\lambda^a\,\lambda^b) = 2\,\delta^{a b}, \qquad
\bigl[ \lambda^a\,,\,\lambda^b \bigr] = 2\,i\,
\mrf^{a b c}\,\lambda_c.
\end{equation}

Finally, there are vertices with no counterpart in the SM, $g g \tbt$ and $g g \mrh \mrh$:

\begin{equation}
\label{nocpo}
\mrV^{\mu\nu\,,\, a b}_{i j} =
\frac{1}{8}\,\frac{g^2_{\mrs} g_6}{\mrM_{\sPW}}\,
\mrf^{c a b}\,\lambda^c_{i j}\,\sigma^{\mu\nu}\,\mra_{t \mrG} ,
\end{equation}

\begin{equation}
\label{nocpt}
\mrV^{\mu\nu\,,\, a b} = \frac{g^2 g_6}{\sqrt{2}\,\mrM^2_{\sPW}}\,
\delta^{a b}\,\bigl( p^{\nu}_3\,p^{\mu}_4 - p_3 \cdot p_4\,\delta^{\mu\nu}
\bigr)\,\mra_{\phi \mrG}.
\end{equation}

In the SESM the only change w.r.t. the SM is a rescaling of the $\tbt \mrh$ coupling by 
$\cos\alpha$.

For the THDM the rescaling is given by

\begin{equation}
\label{THres}
\sin(\alpha - \beta) - \cot \beta\,\cos(\alpha - \beta) 
\end{equation} .

\subsection{\texorpdfstring{$\mrh \to \bbb$}{h to bb}}

Some interesting feature appears when we consider the process $\mrh \to \bbb$ in the SESM.
Beyond LO we will have SM{-}like diagrams rescaled by $\cos\alpha$, producing $\mcO(1/\Lambda^2)$
contributions with normal{-}threshold logarithms for an off{-}shell decaying light Higgs. But there is more: loop 
diagrams with internal $\mrH\,$-lines. Here the expansion has two relevant parameters, $\mrM_{\sPW}/\mrM_s$ and 
$\lambda_{12}$ so that we will have mixed heavy{-}light corrections which are not $\mrM_s\,$-suppressed 
but $\lambda^2_{12}\,$-suppressed, i.e. the amplitude for these diagrams starts with
\begin{equation}
i\,\frac{g^3}{8\,\pi^2}\,\frac{M^3_{b}}{\mrM_{\mrW} \mrM^2_{\mrh}}\,\frac{\mrt^2_3}{\mrt_1} ,
\end{equation} 

\noindent
where the $\mrt_{i}$ are defined in \eqn{tdef}.

\subsection{THDM examples}

An example of complete calculations done in a BSM model, their low{-}energy limit and the corresponding, generalized, kappa{-}framework is as follows:
consider $\mrh \to \gamma \gamma$ in THDM type I. There are two doublets containing fields $\mrh_1, \mrh_2$ which mix with an angle $\beta$;
diagonalization in the neutral sector requires an angle $\alpha$. The amplitude becomes

\begin{equation}
\label{typeI}
\begin{split}
\mrA_{\mrh \gamma \gamma}(s) &=
i\,\frac{g^2\,s^2_{\sPW}}{8\,\pi^2}\,\bigl(
p_1\,\cdot\,p_2\,\delta^{\mu\nu} -
p^{\mu}_2\,p^{\nu}_1 \bigr)
\\
{} &\times \Bigl\{
\frac{\cos\alpha}{\sin\beta}\,\sum_f\,\mrA^{\mySM}_f -
\sin(\alpha - \beta)\,\mrA^{\mySM}_{\sPW}
\\
{} &+ \Bigl[
(s + \mrM^2_{sb})\,\cos(\alpha - \beta)\,\cos(2\,\beta) -
(s + 2\,\mrM^2_{sb} + 2\,\mrM^2_{\mrH^+})\,\sin(\alpha - \beta)\,
\sin(2\,\beta) \Bigr]\,\mrA_{\mrH^+} \Bigr\},
\end{split}
\end{equation}

\noindent
where $\mrM_{sb}$ is the $\mrZ_2$ soft{-}breaking scale~\cite{Yagyu:2012qp,Boggia:2016asg} and the
$\mrh\,$-virtuality is $s$. The coefficients in front of the SM sub{-}amplitudes are kappas and $\mrA_{\mrH^+}$ is the resolved
$\mrH^{+}\,$-loop which will become the contact term in the low{-}energy expansion, $\mrM_{\mrH^+} \to \infty$.

As far as the low{-}energy limit is concerned, one of the possible scenarios is as follows: there are two doublets and the scalar potential will contain a term

\begin{equation}
\label{selectLTHDM}
\mu^2_3\,\bigl( \Phi^{\dagger}_1\,\Phi_2 + \Phi^{\dagger}_2\,\Phi_1 \bigr).
\end{equation}

The scale $\Lambda$ is defined by $\mu^2_3 = \sin\beta\,\cos\beta\,\Lambda^2$, so that all heavy masses are proportional to $\Lambda$ with subleading corrections.
If we denote the VEVs by $\mrv_i$ and introduce $\mrv^2 = \mrv^2 _1 + \mrv^2_2$ the heavy $\Lambda$ limit will give

\begin{equation}
\label{THexp}
\alpha = - \frac{1}{2}\,\frac{\mrv^2}{\Lambda^2} + \mcO(\frac{\mrv^4}{\Lambda^4}) , \qquad
\beta = \frac{1}{2}\,\bigl[ \pi - \frac{\mrv^2}{\Lambda^2} +
\mcO(\frac{\mrv^4}{\Lambda^4}) \Bigr].
\end{equation}

\section{Conclusions \label{conc}}

The price one has to pay for using an EFT in going beyond the SM is that EFTs are only valid in a limited domain.
This prompts the important question as to whether there is a last fundamental theory in this tower of EFTs, each one 
superseding the previous one with rising energies. 
Should one ultimately expect from physics theories to only be valid as approximations and in a limited 
domain~\footnote{Kuhlmann, Meinard, ``Quantum Field Theory'', The Stanford Encyclopedia of Philosophy (Winter 2018
Edition), Edward N. Zalta (ed.).}?

Mathematics suffers from some of the same inherent difficulties as theoretical physics~\footnote{See 
Peter Woit. ``The state of high-energy particle physics: a view from a neighboring field'', 
US Naval Observatory Colloquium December 6, 2018.}: great successes during the 
20th century and increasing difficulties to do better, as the easier problems get solved.
The conventional vision is: some very different physics occurs at Planck scale, SM is just an effective field theory.
What about the next SM? Are we expecting a new weakly-coupled renormalizable model or a tower of EFTs?
There is an alternative vision, the SM could be close to a fundamental theory; indeed,
the lesson from experiments since $1973$ is that it is extremely difficult to find a flaw in the SM. Perhaps the 
SM includes elements of a truly fundamental theory.

Returning to the conventional vision~\cite{Hartmann:2001zz,Bain2013-BAIEFT} we can say that a key ingredient of 
top{-}down EFT studies is matching a given UV theory onto its low{-}energy EFT.
After the Higgs boson discovery, we have a paradigm shift, i.e.\ we use the SMEFT in fits to the data.
The ``fitted'' Wilson coefficients (in the Warsaw basis) can be used to derive SMEFT{-}predicted observables which become 
the pseudo{-}data and we may take any specific BSM model, compute the corresponding low{-}energy limit and confront the 
BSM parameters with the pseudo{-}measurements.

Of course, there is another scenario: depending on the results for the fits, the corresponding interpretation could tell 
us that the required Wilson coefficients are too large to allow for a meaningful interpretation in terms of a weakly 
coupled UV completion. It is also possible that part of the observable parameter (Wilson coefficients) space will be
inconsistent with causality and analyticity.

\acknowledgments

AD gratefully acknowledges the hospitality of the Dipartimento di Fisica in Torino while preparing this work.

\bibliography{reuse2}

\providecommand{\href}[2]{#2}\begingroup\raggedright\begin{thebibliography}{10}

\bibitem{Passarino:2016pzb}
G.~Passarino and M.~Trott, \emph{{The Standard Model Effective Field Theory and
  Next to Leading Order}},  \href{https://arxiv.org/abs/1610.08356}{{\ttfamily
  1610.08356}}.

\bibitem{Brivio:2017vri}
I.~Brivio and M.~Trott, \emph{{The Standard Model as an Effective Field
  Theory}}, \href{https://doi.org/10.1016/j.physrep.2018.11.002}{\emph{Phys.
  Rept.} {\bfseries 793} (2019) 1}
  [\href{https://arxiv.org/abs/1706.08945}{{\ttfamily 1706.08945}}].

\bibitem{Passarino:2019yjx}
G.~Passarino, \emph{{XEFT, the challenging path up the hill: dim = 6 and dim =
  8}},  \href{https://arxiv.org/abs/1901.04177}{{\ttfamily 1901.04177}}.

\bibitem{1798909}
A.~Banerjee and G.~Bhattacharyya, \emph{{Probing the Higgs boson through Yukawa
  force}},  \href{https://arxiv.org/abs/2006.01164}{{\ttfamily 2006.01164}}.

\bibitem{Sirunyan:2018hoz}
{\scshape CMS} collaboration, \emph{{Observation of $\mathrm{t\overline{t}}$H
  production}},
  \href{https://doi.org/10.1103/PhysRevLett.120.231801}{\emph{Phys. Rev. Lett.}
  {\bfseries 120} (2018) 231801}
  [\href{https://arxiv.org/abs/1804.02610}{{\ttfamily 1804.02610}}].

\bibitem{Sirunyan:2018kst}
{\scshape CMS} collaboration, \emph{{Observation of Higgs boson decay to bottom
  quarks}}, \href{https://doi.org/10.1103/PhysRevLett.121.121801}{\emph{Phys.
  Rev. Lett.} {\bfseries 121} (2018) 121801}
  [\href{https://arxiv.org/abs/1808.08242}{{\ttfamily 1808.08242}}].

\bibitem{Aaboud:2018urx}
{\scshape ATLAS} collaboration, \emph{{Observation of Higgs boson production in
  association with a top quark pair at the LHC with the ATLAS detector}},
  \href{https://doi.org/10.1016/j.physletb.2018.07.035}{\emph{Phys. Lett. B}
  {\bfseries 784} (2018) 173}
  [\href{https://arxiv.org/abs/1806.00425}{{\ttfamily 1806.00425}}].

\bibitem{Aaboud:2018zhk}
{\scshape ATLAS} collaboration, \emph{{Observation of $H \rightarrow b\bar{b}$
  decays and $VH$ production with the ATLAS detector}},
  \href{https://doi.org/10.1016/j.physletb.2018.09.013}{\emph{Phys. Lett. B}
  {\bfseries 786} (2018) 59}
  [\href{https://arxiv.org/abs/1808.08238}{{\ttfamily 1808.08238}}].

\bibitem{Brivio:2019irc}
I.~Brivio et~al., \emph{{Computing Tools for the SMEFT}},  in \emph{{Computing
  Tools for the SMEFT}}, J.~Aebischer, M.~Fael, A.~Lenz, M.~Spannowsky and
  J.~Virto, eds., 2019 [\href{https://arxiv.org/abs/1910.11003}{{\ttfamily
  1910.11003}}].

\bibitem{ADtalk}
A.~David and G.~Passarino, \emph{{SMEFT bookkeeping}},  in \emph{LHC EFT
  Working Group: preliminary open discussion}, 2020,
  \href{https://indico.cern.ch/event/908975/}{https://indico.cern.ch/event/908975/}.

\bibitem{Chalons:2016jeu}
G.~Chalons, D.~Lopez-Val, T.~Robens and T.~Stefaniak, \emph{{The Higgs singlet
  extension at LHC Run 2}},
  \href{https://doi.org/10.22323/1.282.1180}{\emph{PoS} {\bfseries ICHEP2016}
  (2016) 1180} [\href{https://arxiv.org/abs/1611.03007}{{\ttfamily
  1611.03007}}].

\bibitem{Yagyu:2012qp}
K.~Yagyu, \emph{{Studies on Extended Higgs Sectors as a Probe of New Physics
  Beyond the Standard Model}}, Ph.D. thesis, Toyama U., 2012.
\newblock \href{https://arxiv.org/abs/1204.0424}{{\ttfamily 1204.0424}}.

\bibitem{Altarelli:2013aqa}
G.~Altarelli and D.~Meloni, \emph{{A non supersymmetric SO(10) grand unified
  model for all the physics below $M_{GUT}$}},
  \href{https://doi.org/10.1007/JHEP08(2013)021}{\emph{JHEP} {\bfseries 08}
  (2013) 021} [\href{https://arxiv.org/abs/1305.1001}{{\ttfamily 1305.1001}}].

\bibitem{Okada:2016whh}
H.~Okada, N.~Okada, Y.~Orikasa and K.~Yagyu, \emph{{Higgs phenomenology in the
  minimal SU(3)$_L\times$U(1)$_X$ model}},
  \href{https://doi.org/10.1103/PhysRevD.94.015002}{\emph{Phys. Rev. D}
  {\bfseries 94} (2016) 015002}
  [\href{https://arxiv.org/abs/1604.01948}{{\ttfamily 1604.01948}}].

\bibitem{Brivio:2013pma}
I.~Brivio, T.~Corbett, O.J.P.~Éboli, M.B.~Gavela, J.~Gonzalez-Fraile,
  M.C.~Gonzalez-Garcia et~al., \emph{{Disentangling a dynamical Higgs}},
  \href{https://doi.org/10.1007/JHEP03(2014)024}{\emph{JHEP} {\bfseries 03}
  (2014) 024} [\href{https://arxiv.org/abs/1311.1823}{{\ttfamily 1311.1823}}].

\bibitem{Buchalla:2015qju}
G.~Buchalla, O.~Cata, A.~Celis and C.~Krause, \emph{{Fitting Higgs Data with
  Nonlinear Effective Theory}},
  \href{https://doi.org/10.1140/epjc/s10052-016-4086-9}{\emph{Eur. Phys. J. C}
  {\bfseries 76} (2016) 233}
  [\href{https://arxiv.org/abs/1511.00988}{{\ttfamily 1511.00988}}].

\bibitem{Cohen:2020xca}
T.~Cohen, N.~Craig, X.~Lu and D.~Sutherland, \emph{{Is SMEFT Enough?}},
  \href{https://arxiv.org/abs/2008.08597}{{\ttfamily 2008.08597}}.

\bibitem{Einhorn:2013kja}
M.B.~Einhorn and J.~Wudka, \emph{{The Bases of Effective Field Theories}},
  \href{https://doi.org/10.1016/j.nuclphysb.2013.08.023}{\emph{Nucl. Phys. B}
  {\bfseries 876} (2013) 556}
  [\href{https://arxiv.org/abs/1307.0478}{{\ttfamily 1307.0478}}].

\bibitem{Lee:1977qs}
B.W.~Lee and S.~Weinberg, \emph{{SU(3) x U(1) Gauge Theory of the Weak and
  Electromagnetic Interactions}},
  \href{https://doi.org/10.1103/PhysRevLett.38.1237}{\emph{Phys. Rev. Lett.}
  {\bfseries 38} (1977) 1237}.

\bibitem{Hays:2020scx}
C.~Hays, A.~Helset, A.~Martin and M.~Trott, \emph{{Exact SMEFT formulation and
  expansion to $\mathcal{O}(v^4/\Lambda^4)$}},
  \href{https://arxiv.org/abs/2007.00565}{{\ttfamily 2007.00565}}.

\bibitem{Grzadkowski:2010es}
B.~Grzadkowski, M.~Iskrzynski, M.~Misiak and J.~Rosiek, \emph{{Dimension-Six
  Terms in the Standard Model Lagrangian}},
  \href{https://doi.org/10.1007/JHEP10(2010)085}{\emph{JHEP} {\bfseries 10}
  (2010) 085} [\href{https://arxiv.org/abs/1008.4884}{{\ttfamily 1008.4884}}].

\bibitem{Arzt:1993gz}
C.~Arzt, \emph{{Reduced effective Lagrangians}},
  \href{https://doi.org/10.1016/0370-2693(94)01419-D}{\emph{Phys. Lett. B}
  {\bfseries 342} (1995) 189}
  [\href{https://arxiv.org/abs/hep-ph/9304230}{{\ttfamily hep-ph/9304230}}].

\bibitem{Ghezzi:2015vva}
M.~Ghezzi, R.~Gomez-Ambrosio, G.~Passarino and S.~Uccirati, \emph{{NLO Higgs
  effective field theory and $\kappa$-framework}},
  \href{https://doi.org/10.1007/JHEP07(2015)175}{\emph{JHEP} {\bfseries 07}
  (2015) 175} [\href{https://arxiv.org/abs/1505.03706}{{\ttfamily
  1505.03706}}].

\bibitem{Wells:2016luz}
J.D.~Wells, \emph{{Higgs naturalness and the scalar boson proliferation
  instability problem}},
  \href{https://doi.org/10.1007/s11229-014-0618-8}{\emph{Synthese} {\bfseries
  194} (2017) 477} [\href{https://arxiv.org/abs/1603.06131}{{\ttfamily
  1603.06131}}].

\bibitem{Cullen:2019nnr}
J.M.~Cullen, B.D.~Pecjak and D.J.~Scott, \emph{{NLO corrections to $h\to b\bar
  b$ decay in SMEFT}},
  \href{https://doi.org/10.1007/JHEP08(2019)173}{\emph{JHEP} {\bfseries 08}
  (2019) 173} [\href{https://arxiv.org/abs/1904.06358}{{\ttfamily
  1904.06358}}].

\bibitem{Cullen:2020zof}
J.M.~Cullen and B.D.~Pecjak, \emph{{Higgs decay to fermion pairs at NLO in
  SMEFT}},  \href{https://arxiv.org/abs/2007.15238}{{\ttfamily 2007.15238}}.

\bibitem{Gauld:2016kuu}
R.~Gauld, B.D.~Pecjak and D.J.~Scott, \emph{{QCD radiative corrections for
  $h\to b\bar b$ in the Standard Model Dimension-6 EFT}},
  \href{https://doi.org/10.1103/PhysRevD.94.074045}{\emph{Phys. Rev. D}
  {\bfseries 94} (2016) 074045}
  [\href{https://arxiv.org/abs/1607.06354}{{\ttfamily 1607.06354}}].

\bibitem{Buchalla:2019wsc}
G.~Buchalla, A.~Celis, C.~Krause and J.-N.~Toelstede, \emph{{Master Formula for
  One-Loop Renormalization of Bosonic SMEFT Operators}},
  \href{https://arxiv.org/abs/1904.07840}{{\ttfamily 1904.07840}}.

\bibitem{tHooft:1973bhk}
G.~'t~Hooft, \emph{{An algorithm for the poles at dimension four in the
  dimensional regularization procedure}},
  \href{https://doi.org/10.1016/0550-3213(73)90263-0}{\emph{Nucl. Phys. B}
  {\bfseries 62} (1973) 444}.

\bibitem{tHooft:1974toh}
G.~'t~Hooft and M.~Veltman, \emph{{One loop divergencies in the theory of
  gravitation}}, {\emph{Ann. Inst. H. Poincare Phys. Theor. A} {\bfseries 20}
  (1974) 69}.

\bibitem{delAguila:2016zcb}
F.~del Aguila, Z.~Kunszt and J.~Santiago, \emph{{One-loop effective lagrangians
  after matching}},
  \href{https://doi.org/10.1140/epjc/s10052-016-4081-1}{\emph{Eur. Phys. J.}
  {\bfseries C76} (2016) 244}
  [\href{https://arxiv.org/abs/1602.00126}{{\ttfamily 1602.00126}}].

\bibitem{Jiang:2018pbd}
M.~Jiang, N.~Craig, Y.-Y.~Li and D.~Sutherland, \emph{{Complete One-Loop
  Matching for a Singlet Scalar in the Standard Model EFT}},
  \href{https://doi.org/10.1007/JHEP02(2019)031}{\emph{JHEP} {\bfseries 02}
  (2019) 031} [\href{https://arxiv.org/abs/1811.08878}{{\ttfamily
  1811.08878}}].

\bibitem{Donoghue:2017pgk}
J.F.~Donoghue, M.M.~Ivanov and A.~Shkerin, \emph{{EPFL Lectures on General
  Relativity as a Quantum Field Theory}},
  \href{https://arxiv.org/abs/1702.00319}{{\ttfamily 1702.00319}}.

\bibitem{Passarino:2018wix}
G.~Passarino, \emph{{Peaks and cusps: anomalous thresholds and LHC physics}},
  \href{https://arxiv.org/abs/1807.00503}{{\ttfamily 1807.00503}}.

\bibitem{pulls}
L.~Demortier and L.~Lyons, ``{{Everything you always wanted to know about
  pulls}}.''
  \href{https://lucdemortier.github.io/assets/papers/cdf5776_pulls.pdf}{CDF/ANAL/PPUBLIC/5776},
  2008.

\bibitem{vanBeek:2019evb}
S.~van Beek, E.R.~Nocera, J.~Rojo and E.~Slade, \emph{{Constraining the SMEFT
  with Bayesian reweighting}},
  \href{https://doi.org/10.21468/SciPostPhys.7.5.070}{\emph{SciPost Phys.}
  {\bfseries 7} (2019) 070} [\href{https://arxiv.org/abs/1906.05296}{{\ttfamily
  1906.05296}}].

\bibitem{Hays:2018zze}
C.~Hays, A.~Martin, V.~Sanz and J.~Setford, \emph{{On the impact of
  dimension-eight SMEFT operators on Higgs measurements}},
  \href{https://doi.org/10.1007/JHEP02(2019)123}{\emph{JHEP} {\bfseries 02}
  (2019) 123} [\href{https://arxiv.org/abs/1808.00442}{{\ttfamily
  1808.00442}}].

\bibitem{LHCHiggsCrossSectionWorkingGroup:2012nn}
{\scshape LHC Higgs Cross Section Working Group} collaboration, \emph{{LHC
  HXSWG interim recommendations to explore the coupling structure of a
  Higgs-like particle}},  \href{https://arxiv.org/abs/1209.0040}{{\ttfamily
  1209.0040}}.

\bibitem{Passarino:2013nka}
G.~Passarino, \emph{{Higgs Boson Production and Decay: Dalitz Sector}},
  \href{https://doi.org/10.1016/j.physletb.2013.10.052}{\emph{Phys. Lett. B}
  {\bfseries 727} (2013) 424}
  [\href{https://arxiv.org/abs/1308.0422}{{\ttfamily 1308.0422}}].

\bibitem{collaboration2020search}
{\scshape ATLAS} collaboration, \emph{{A search for the $Z\gamma$ decay mode of
  the Higgs boson in $pp$ collisions at $\sqrt{s}$ = 13 TeV with the ATLAS
  detector}},  \href{https://arxiv.org/abs/2005.05382}{{\ttfamily 2005.05382}}.

\bibitem{Kachanovich:2020xyg}
A.~Kachanovich, U.~Nierste and I.~Ni\v{s}and\v{z}i\'c, \emph{{Higgs boson decay
  into a lepton pair and a photon revisited}},
  \href{https://doi.org/10.1103/PhysRevD.101.073003}{\emph{Phys. Rev. D}
  {\bfseries 101} (2020) 073003}
  [\href{https://arxiv.org/abs/2001.06516}{{\ttfamily 2001.06516}}].

\bibitem{Abbasabadi:1996ze}
A.~Abbasabadi, D.~Bowser-Chao, D.A.~Dicus and W.W.~Repko, \emph{{Radiative
  Higgs boson decays H ---> fermion anti-fermion gamma}},
  \href{https://doi.org/10.1103/PhysRevD.55.5647}{\emph{Phys. Rev. D}
  {\bfseries 55} (1997) 5647}
  [\href{https://arxiv.org/abs/hep-ph/9611209}{{\ttfamily hep-ph/9611209}}].

\bibitem{Abbasabadi:2006dd}
A.~Abbasabadi and W.W.~Repko, \emph{{Higgs boson decay into Z bosons and a
  photon}}, \href{https://doi.org/10.1088/1126-6708/2006/08/048}{\emph{JHEP}
  {\bfseries 08} (2006) 048}
  [\href{https://arxiv.org/abs/hep-ph/0602087}{{\ttfamily hep-ph/0602087}}].

\bibitem{Dicus:2013ycd}
D.A.~Dicus and W.W.~Repko, \emph{{Calculation of the decay $H\to
  e\bar{e}\gamma$}},
  \href{https://doi.org/10.1103/PhysRevD.87.077301}{\emph{Phys. Rev. D}
  {\bfseries 87} (2013) 077301}
  [\href{https://arxiv.org/abs/1302.2159}{{\ttfamily 1302.2159}}].

\bibitem{Denner:2018opp}
A.~Denner, S.~Dittmaier and J.-N.~Lang, \emph{{Renormalization of mixing
  angles}}, \href{https://doi.org/10.1007/JHEP11(2018)104}{\emph{JHEP}
  {\bfseries 11} (2018) 104}
  [\href{https://arxiv.org/abs/1808.03466}{{\ttfamily 1808.03466}}].

\bibitem{Low:2012rj}
I.~Low, J.~Lykken and G.~Shaughnessy, \emph{{Have We Observed the Higgs
  (Imposter)?}}, \href{https://doi.org/10.1103/PhysRevD.86.093012}{\emph{Phys.
  Rev.} {\bfseries D86} (2012) 093012}
  [\href{https://arxiv.org/abs/1207.1093}{{\ttfamily 1207.1093}}].

\bibitem{Einhorn:1981cy}
M.B.~Einhorn, D.R.T.~Jones and M.J.G.~Veltman, \emph{{Heavy Particles and the
  rho Parameter in the Standard Model}},
  \href{https://doi.org/10.1016/0550-3213(81)90292-3}{\emph{Nucl. Phys.}
  {\bfseries B191} (1981) 146}.

\bibitem{Passarino:1990nu}
G.~Passarino, \emph{{Radiative corrections to the Rho parameter versus the top
  quark mass}}, \href{https://doi.org/10.1016/0370-2693(90)91906-R}{\emph{Phys.
  Lett.} {\bfseries B247} (1990) 587}.

\bibitem{Lynn:1990zk}
B.W.~Lynn and E.~Nardi, \emph{{Radiative corrections in unconstrained SU(2) x
  U(1) and the top mass problem}},
  \href{https://doi.org/10.1016/0550-3213(92)90486-U}{\emph{Nucl. Phys.}
  {\bfseries B381} (1992) 467}.

\bibitem{Boggia:2016asg}
M.~Boggia, R.~Gomez-Ambrosio and G.~Passarino, \emph{{Low energy behaviour of
  standard model extensions}},
  \href{https://doi.org/10.1007/JHEP05(2016)162}{\emph{JHEP} {\bfseries 05}
  (2016) 162} [\href{https://arxiv.org/abs/1603.03660}{{\ttfamily
  1603.03660}}].

\bibitem{Buchalla:2016bse}
G.~Buchalla, O.~Cata, A.~Celis and C.~Krause, \emph{{Standard Model Extended by
  a Heavy Singlet: Linear vs. Nonlinear EFT}},
  \href{https://doi.org/10.1016/j.nuclphysb.2017.02.006}{\emph{Nucl. Phys. B}
  {\bfseries 917} (2017) 209}
  [\href{https://arxiv.org/abs/1608.03564}{{\ttfamily 1608.03564}}].

\bibitem{Jenkins:2017jig}
E.E.~Jenkins, A.V.~Manohar and P.~Stoffer, \emph{{Low-Energy Effective Field
  Theory below the Electroweak Scale: Operators and Matching}},
  \href{https://doi.org/10.1007/JHEP03(2018)016}{\emph{JHEP} {\bfseries 03}
  (2018) 016} [\href{https://arxiv.org/abs/1709.04486}{{\ttfamily
  1709.04486}}].

\bibitem{Gorbahn:2015gxa}
M.~Gorbahn, J.M.~No and V.~Sanz, \emph{{Benchmarks for Higgs Effective Theory:
  Extended Higgs Sectors}},
  \href{https://doi.org/10.1007/JHEP10(2015)036}{\emph{JHEP} {\bfseries 10}
  (2015) 036} [\href{https://arxiv.org/abs/1502.07352}{{\ttfamily
  1502.07352}}].

\bibitem{Brehmer:2015rna}
J.~Brehmer, A.~Freitas, D.~Lopez-Val and T.~Plehn, \emph{{Pushing Higgs
  Effective Theory to its Limits}},
  \href{https://doi.org/10.1103/PhysRevD.93.075014}{\emph{Phys. Rev. D}
  {\bfseries 93} (2016) 075014}
  [\href{https://arxiv.org/abs/1510.03443}{{\ttfamily 1510.03443}}].

\bibitem{Walker:2013hka}
D.G.E.~Walker, \emph{{Unitarity Constraints on Higgs Portals}},
  \href{https://arxiv.org/abs/1310.1083}{{\ttfamily 1310.1083}}.

\bibitem{Appelquist:1974tg}
T.~Appelquist and J.~Carazzone, \emph{{Infrared Singularities and Massive
  Fields}}, \href{https://doi.org/10.1103/PhysRevD.11.2856}{\emph{Phys. Rev. D}
  {\bfseries 11} (1975) 2856}.

\bibitem{Robens:2015gla}
T.~Robens and T.~Stefaniak, \emph{{Status of the Higgs Singlet Extension of the
  Standard Model after LHC Run 1}},
  \href{https://doi.org/10.1140/epjc/s10052-015-3323-y}{\emph{Eur. Phys. J. C}
  {\bfseries 75} (2015) 104}
  [\href{https://arxiv.org/abs/1501.02234}{{\ttfamily 1501.02234}}].

\bibitem{PhysRevD.88.115012}
G.M.~Pruna and T.~Robens, \emph{{Higgs singlet extension parameter space in the
  light of the LHC discovery}},
  \href{https://doi.org/10.1103/PhysRevD.88.115012}{\emph{Phys. Rev. D}
  {\bfseries 88} (2013) 115012}.

\bibitem{Veltman:1989vw}
M.~Veltman and F.~Yndurain, \emph{{Radiative corrections to $\mrW \mrW$
  scattering}}, \href{https://doi.org/10.1016/0550-3213(89)90369-6}{\emph{Nucl.
  Phys. B} {\bfseries 325} (1989) 1}.

\bibitem{Biekotter:2016ecg}
A.~Biekötter, J.~Brehmer and T.~Plehn, \emph{{Extending the limits of Higgs
  effective theory}},
  \href{https://doi.org/10.1103/PhysRevD.94.055032}{\emph{Phys. Rev. D}
  {\bfseries 94} (2016) 055032}
  [\href{https://arxiv.org/abs/1602.05202}{{\ttfamily 1602.05202}}].

\bibitem{Osipov:2001th}
A.A.~Osipov and B.~Hiller, \emph{{Inverse mass expansion of the one-loop
  effective action}},
  \href{https://doi.org/10.1016/S0370-2693(01)00889-9}{\emph{Phys. Lett.}
  {\bfseries B515} (2001) 458}
  [\href{https://arxiv.org/abs/hep-th/0104165}{{\ttfamily hep-th/0104165}}].

\bibitem{Altenkamp:2018bcs}
L.~Altenkamp, M.~Boggia and S.~Dittmaier, \emph{{Precision calculations for $h
  \to WW/ZZ \to 4$ fermions in a Singlet Extension of the Standard Model with
  Prophecy4f}}, \href{https://doi.org/10.1007/JHEP04(2018)062}{\emph{JHEP}
  {\bfseries 04} (2018) 062}
  [\href{https://arxiv.org/abs/1801.07291}{{\ttfamily 1801.07291}}].

\bibitem{Kanemura:2016lkz}
S.~Kanemura, M.~Kikuchi and K.~Yagyu, \emph{{One-loop corrections to the Higgs
  self-couplings in the singlet extension}},
  \href{https://doi.org/10.1016/j.nuclphysb.2017.02.004}{\emph{Nucl. Phys. B}
  {\bfseries 917} (2017) 154}
  [\href{https://arxiv.org/abs/1608.01582}{{\ttfamily 1608.01582}}].

\bibitem{Boughezal:2020uwq}
R.~Boughezal, F.~Petriello and D.~Wiegand, \emph{{Removing flat directions in
  standard model EFT fits: How polarized electron-ion collider data can
  complement the LHC}},
  \href{https://doi.org/10.1103/PhysRevD.101.116002}{\emph{Phys. Rev. D}
  {\bfseries 101} (2020) 116002}
  [\href{https://arxiv.org/abs/2004.00748}{{\ttfamily 2004.00748}}].

\bibitem{Brehmer:2016nyr}
J.~Brehmer, K.~Cranmer, F.~Kling and T.~Plehn, \emph{{Better Higgs boson
  measurements through information geometry}},
  \href{https://doi.org/10.1103/PhysRevD.95.073002}{\emph{Phys. Rev. D}
  {\bfseries 95} (2017) 073002}
  [\href{https://arxiv.org/abs/1612.05261}{{\ttfamily 1612.05261}}].

\bibitem{MIG}
S.~Amari and H.~Nagaoka, ``{Methods of Information Geometry}.'' {volume 191 of
  Translations of Mathematical Monographs. American Mathematical Society,
  2000}, 2000.

\bibitem{Bodwin:2019ivc}
G.T.~Bodwin and H.S.~Chung, \emph{{New method for fitting coefficients in
  standard model effective theory}},
  \href{https://doi.org/10.1103/PhysRevD.101.115039}{\emph{Phys. Rev. D}
  {\bfseries 101} (2020) 115039}
  [\href{https://arxiv.org/abs/1912.09843}{{\ttfamily 1912.09843}}].

\bibitem{TT}
M.K.~Transtrum, B.B.~Machta, K.S.~Brown, B.C.~Daniels, C.R.~Myers and
  J.P.~Sethna, \emph{Perspective: Sloppiness and emergent theories in physics,
  biology, and beyond}, \href{https://doi.org/10.1063/1.4923066}{\emph{The
  Journal of Chemical Physics} {\bfseries 143} (2015) 010901}.

\bibitem{Kalinowski:2018oxd}
J.~Kalinowski, P.~Kozów, S.~Pokorski, J.~Rosiek, M.~Szleper and S.a.~Tkaczyk,
  \emph{{Same-sign WW scattering at the LHC: can we discover BSM effects before
  discovering new states?}},
  \href{https://doi.org/10.1140/epjc/s10052-018-5885-y}{\emph{Eur. Phys. J. C}
  {\bfseries 78} (2018) 403}
  [\href{https://arxiv.org/abs/1802.02366}{{\ttfamily 1802.02366}}].

\bibitem{Bhatta}
A.~Bhattacharyya, ``{{On a measure of divergence between two statistical
  populations defined by probability distributions }}.'' {Bull. Calcutta Math.
  Soc. , 35 (1943) pp. 99–109}, 1943.

\bibitem{shemyakin2014}
A.~Shemyakin, \emph{Hellinger distance and non-informative priors},
  \href{https://doi.org/10.1214/14-BA881}{\emph{Bayesian Anal.} {\bfseries 9}
  (2014) 923}.

\bibitem{Brehmer:2017lrt}
J.~Brehmer, F.~Kling, T.~Plehn and T.M.P.~Tait, \emph{{Better Higgs-CP Tests
  Through Information Geometry}},
  \href{https://doi.org/10.1103/PhysRevD.97.095017}{\emph{Phys. Rev. D}
  {\bfseries 97} (2018) 095017}
  [\href{https://arxiv.org/abs/1712.02350}{{\ttfamily 1712.02350}}].

\bibitem{TamBoo}
R.~Tamura and D.~Boos, ``{{Minimum Hellinger Distance Estimation for
  Multivariate Location and Covariance}}.'' {Journal of the American
  Statistical Association, 81, 223-229}, 1986.

\bibitem{Alvarez:2019knh}
E.~Alvarez, F.~Lamagna and M.~Szewc, \emph{{Topic Model for four-top at the
  LHC}}, \href{https://doi.org/10.1007/JHEP01(2020)049}{\emph{JHEP} {\bfseries
  20} (2020) 049} [\href{https://arxiv.org/abs/1911.09699}{{\ttfamily
  1911.09699}}].

\bibitem{LOURENZUTTI20144414}
R.~Lourenzutti and R.A.~Krohling, \emph{{The Hellinger distance in
  Multicriteria Decision Making: An illustration to the TOPSIS and TODIM
  methods}},
  \href{https://doi.org/https://doi.org/10.1016/j.eswa.2014.01.015}{\emph{Expert
  Systems with Applications} {\bfseries 41} (2014) 4414 }.

\bibitem{carter2008information}
K.M.~Carter, R.~Raich and A.O.H.~III, \emph{{An Information Geometric Framework
  for Dimensionality Reduction}},
  \href{https://arxiv.org/abs/0809.4866}{{\ttfamily 0809.4866}}.

\bibitem{10.2307/2289852}
D.G.~Simpson, \emph{Hellinger deviance tests: Efficiency, breakdown points, and
  examples}, {\emph{Journal of the American Statistical Association} {\bfseries
  84} (1989) 107}.

\bibitem{canonne2018structure}
C.L.~Canonne, G.~Kamath, A.~McMillan, A.~Smith and J.~Ullman, \emph{{The
  Structure of Optimal Private Tests for Simple Hypotheses}},
  \href{https://arxiv.org/abs/1811.11148}{{\ttfamily 1811.11148}}.

\bibitem{Krauss:2017xpj}
M.E.~Krauss and F.~Staub, \emph{{Perturbativity Constraints in BSM Models}},
  \href{https://doi.org/10.1140/epjc/s10052-018-5676-5}{\emph{Eur. Phys. J. C}
  {\bfseries 78} (2018) 185}
  [\href{https://arxiv.org/abs/1709.03501}{{\ttfamily 1709.03501}}].

\bibitem{Wells:2017aoy}
J.D.~Wells, Z.~Zhang and Y.~Zhao, \emph{{Establishing the Isolated Standard
  Model}}, \href{https://doi.org/10.1103/PhysRevD.96.015005}{\emph{Phys. Rev.}
  {\bfseries D96} (2017) 015005}
  [\href{https://arxiv.org/abs/1702.06954}{{\ttfamily 1702.06954}}].

\bibitem{Hartland:2019bjb}
N.P.~Hartland, F.~Maltoni, E.R.~Nocera, J.~Rojo, E.~Slade, E.~Vryonidou et~al.,
  \emph{{A Monte Carlo global analysis of the Standard Model Effective Field
  Theory: the top quark sector}},
  \href{https://doi.org/10.1007/JHEP04(2019)100}{\emph{JHEP} {\bfseries 04}
  (2019) 100} [\href{https://arxiv.org/abs/1901.05965}{{\ttfamily
  1901.05965}}].

\bibitem{Ahn:1988fx}
C.~Ahn, M.E.~Peskin, B.W.~Lynn and S.B.~Selipsky, \emph{{Delayed Unitarity
  Cancellation and Heavy Particle Effects in $e^+ e^- \to W^+ W^-$}},
  \href{https://doi.org/10.1016/0550-3213(88)90081-8}{\emph{Nucl. Phys.}
  {\bfseries B309} (1988) 221}.

\bibitem{Kallosh:1972ap}
R.~Kallosh and I.~Tyutin, \emph{{The Equivalence theorem and gauge invariance
  in renormalizable theories}}, {\emph{Yad. Fiz.} {\bfseries 17} (1973) 190}.

\bibitem{Denner:2019fcr}
A.~Denner, S.~Dittmaier and A.~Mück, \emph{{PROPHECY4F 3.0: A Monte Carlo
  program for Higgs-boson decays into four-fermion final states in and beyond
  the Standard Model}},
  \href{https://doi.org/10.1016/j.cpc.2020.107336}{\emph{Comput. Phys. Commun.}
  {\bfseries 254} (2020) 107336}
  [\href{https://arxiv.org/abs/1912.02010}{{\ttfamily 1912.02010}}].

\bibitem{Brivio:2019myy}
I.~Brivio, T.~Corbett and M.~Trott, \emph{{The Higgs width in the SMEFT}},
  \href{https://doi.org/10.1007/JHEP10(2019)056}{\emph{JHEP} {\bfseries 10}
  (2019) 056} [\href{https://arxiv.org/abs/1906.06949}{{\ttfamily
  1906.06949}}].

\bibitem{Adams:2006sv}
A.~Adams, N.~Arkani-Hamed, S.~Dubovsky, A.~Nicolis and R.~Rattazzi,
  \emph{{Causality, analyticity and an IR obstruction to UV completion}},
  \href{https://doi.org/10.1088/1126-6708/2006/10/014}{\emph{JHEP} {\bfseries
  10} (2006) 014} [\href{https://arxiv.org/abs/hep-th/0602178}{{\ttfamily
  hep-th/0602178}}].

\bibitem{Zhang:2018shp}
C.~Zhang and S.-Y.~Zhou, \emph{{Positivity bounds on vector boson scattering at
  the LHC}}, \href{https://doi.org/10.1103/PhysRevD.100.095003}{\emph{Phys.
  Rev. D} {\bfseries 100} (2019) 095003}
  [\href{https://arxiv.org/abs/1808.00010}{{\ttfamily 1808.00010}}].

\bibitem{Bardin:1999gt}
D.Y.~Bardin, M.~Grunewald and G.~Passarino, \emph{{Precision calculation
  project report}},  \href{https://arxiv.org/abs/hep-ph/9902452}{{\ttfamily
  hep-ph/9902452}}.

\bibitem{Helset:2020yio}
A.~Helset, A.~Martin and M.~Trott, \emph{{The Geometric Standard Model
  Effective Field Theory}},
  \href{https://doi.org/10.1007/JHEP03(2020)163}{\emph{JHEP} {\bfseries 03}
  (2020) 163} [\href{https://arxiv.org/abs/2001.01453}{{\ttfamily
  2001.01453}}].

\bibitem{Murphy:2020rsh}
C.W.~Murphy, \emph{{Dimension-8 Operators in the Standard Model Effective Field
  Theory}},  \href{https://arxiv.org/abs/2005.00059}{{\ttfamily 2005.00059}}.

\bibitem{Li:2020gnx}
H.-L.~Li, Z.~Ren, J.~Shu, M.-L.~Xiao, J.-H.~Yu and Y.-H.~Zheng, \emph{{Complete
  Set of Dimension-8 Operators in the Standard Model Effective Field Theory}},
  \href{https://arxiv.org/abs/2005.00008}{{\ttfamily 2005.00008}}.

\bibitem{Criado:2018sdb}
J.~Criado and M.~Pérez-Victoria, \emph{{Field redefinitions in effective
  theories at higher orders}},
  \href{https://doi.org/10.1007/JHEP03(2019)038}{\emph{JHEP} {\bfseries 03}
  (2019) 038} [\href{https://arxiv.org/abs/1811.09413}{{\ttfamily
  1811.09413}}].

\bibitem{Passarino:2016saj}
G.~Passarino, \emph{{Field reparametrization in effective field theories}},
  \href{https://doi.org/10.1140/epjp/i2017-11291-5}{\emph{Eur. Phys. J. Plus}
  {\bfseries 132} (2017) 16}
  [\href{https://arxiv.org/abs/1610.09618}{{\ttfamily 1610.09618}}].

\bibitem{Goria:2011wa}
S.~Goria, G.~Passarino and D.~Rosco, \emph{{The Higgs Boson Lineshape}},
  \href{https://doi.org/10.1016/j.nuclphysb.2012.07.006}{\emph{Nucl. Phys. B}
  {\bfseries 864} (2012) 530}
  [\href{https://arxiv.org/abs/1112.5517}{{\ttfamily 1112.5517}}].

\bibitem{Bojarski:2015kra}
F.~Bojarski, G.~Chalons, D.~Lopez-Val and T.~Robens, \emph{{Heavy to light
  Higgs boson decays at NLO in the Singlet Extension of the Standard Model}},
  \href{https://doi.org/10.1007/JHEP02(2016)147}{\emph{JHEP} {\bfseries 02}
  (2016) 147} [\href{https://arxiv.org/abs/1511.08120}{{\ttfamily
  1511.08120}}].

\bibitem{Fuentes-Martin:2016uol}
J.~Fuentes-Martin, J.~Portoles and P.~Ruiz-Femenia, \emph{{Integrating out
  heavy particles with functional methods: a simplified framework}},
  \href{https://doi.org/10.1007/JHEP09(2016)156}{\emph{JHEP} {\bfseries 09}
  (2016) 156} [\href{https://arxiv.org/abs/1607.02142}{{\ttfamily
  1607.02142}}].

\bibitem{vanderBij:1983bw}
J.~van~der Bij and M.~Veltman, \emph{{Two Loop Large Higgs Mass Correction to
  the rho Parameter}},
  \href{https://doi.org/10.1016/0550-3213(84)90284-0}{\emph{Nucl. Phys. B}
  {\bfseries 231} (1984) 205}.

\bibitem{Hartmann:2001zz}
S.~Hartmann, \emph{{Effective field theories, reductionism and scientific
  explanation}},
  \href{https://doi.org/10.1016/S1355-2198(01)00005-3}{\emph{Stud. Hist.
  Philos. Mod. Phys.} {\bfseries 32} (2001) 267}.

\bibitem{Bain2013-BAIEFT}
J.~Bain, \emph{{Effective Field Theories}},  in \emph{The Oxford Handbook of
  Philosophy of Physics}, R.~Batterman, ed., p.~224, Oxford University Press
  (2013), \href{https://doi.org/10.1093/oxfordhb/9780195392043.001.0001}{DOI}.

\end{thebibliography}\endgroup
\bibliographystyle{JHEP}

\end{document}